\documentclass[useAMS]{mn2e}

\usepackage{graphicx}

\usepackage{color}

\def\gsim{\mathrel{\rlap{\lower 4pt \hbox{\hskip 1pt $\sim$}}\raise 1pt
\hbox {$>$}}}
\def\lsim{\mathrel{\rlap{\lower 4pt \hbox{\hskip 1pt $\sim$}}\raise 1pt
\hbox {$<$}}}

\title[Explosion Asymmetry in Supernova Color and Luminosity]
{Effects of Explosion Asymmetry and Viewing Angle on 
the Type Ia Supernova Color and Luminosity Calibration\thanks{
Partially based on observations obtained at the Gemini Observatory,
Cerro Pachon, Chile (Gemini Programs GS-2008B-Q-8, GS-2008B-Q-32, 
GS-2008B-Q-56, and GS-2009B-Q-40).}}
\author[K. Maeda et al.]
{Keiichi Maeda$^{1}$, 
Giorgos Leloudas$^{2}$, 
Stefan Taubenberger$^{3}$, 
\newauthor
Maximilian Stritzinger$^{4,5,2}$, 
Jesper Sollerman$^{4,2}$, 
Nancy Elias-Rosa$^{6}$, 
\newauthor
Stefano Benetti$^{7}$, 
Mario Hamuy$^{8}$, 
Gaston Folatelli$^{8,1}$, 
Paolo A. Mazzali$^{3,9}$\\
$^{1}$Institute for the Physics and Mathematics of the Universe (IPMU), 
Todai Institutes for Advanced Study (TODIAS), \\
University of Tokyo, 5-1-5 Kashiwanoha, 
Kashiwa, Chiba 277-8583, Japan; keiichi.maeda@ipmu.jp\\
$^{2}$Dark Cosmology Centre, Niels Bohr Institute, Copenhagen University, 
Juliane Maries Vej 30, 2100 Copenhagen \O, Denmark\\
$^{3}$Max-Planck-Institut f\"ur Astrophysik, 
Karl-Schwarzschild-Stra{\ss}e 1, 85741 Garching, Germany\\
$^{4}$The Oskar Klein Centre, Department of Astronomy, 
Stockholm University, AlbaNova, 10691 Stockholm, Sweden\\
$^{5}$Carnegie Observatories, Las Campanas Observatory, Casilla 601, La Serena, Chile\\
$^{6}$Spitzer Science Center, California Institute of Technology, 1200 E. 
California Blvd., Pasadena, CA 91125, USA\\
$^{7}$INAF - Osservatorio Astronomico di Padova, vicolo dell'Osservatorio 5, I-35122 Padova, Italy\\
$^{8}$Departamento de Astronom\'ia, Universidad de Chile, Casilla 36-D, Santiago, Chile\\
$^{9}$Scuola Normale Superiore, Piazza Cavalieri 7, 56127 Pisa, Italy\\
}
\begin{document}

\date{}

\pagerange{\pageref{firstpage}--\pageref{lastpage}} \pubyear{2010}

\maketitle

\label{firstpage}

\begin{abstract}
Phenomenological relations exist between the peak luminosity and other 
observables of type Ia supernovae (SNe~Ia), that allow one to standardize their peak luminosities. 
However, several issues are yet to be clarified: SNe~Ia show color variations 
after the standardization. Also, individual SNe~Ia can show residuals in their standardized 
peak absolute magnitude at the level of  $\sim 0.15$ mag. 
In this paper, we explore how the color and luminosity residual are related to 
the wavelength shift of nebular emission lines observed at $\gsim 150$ 
days after maximum light. 
A sample of 11 SNe Ia which likely suffer from little host extinction indicates 
a correlation ($3.3\sigma$) between the peak $B-V$ color 
and the late-time emission-line shift. 
Furthermore, a nearly identical relation applies for a larger sample in which 
only three SNe with $B-V \gsim 0.2$ mag are excluded.
Following the interpretation that the late-time emission-line shift 
is a tracer of the viewing direction from which an off-centre explosion is observed, 
we suggest that the viewing direction is a dominant factor controlling the 
SN color and that a large part of the color variations is intrinsic, 
rather than due to the host extinction. 
We also investigate a relation between the peak luminosity residuals 
and the wavelength shift in nebular emission lines in a sample of 
20 SNe. 
We thereby found a hint of a correlation (at $\sim 1.6 \sigma$ level). 
The confirmation of this will require a future sample of SNe with more 
accurate distance estimates.
Radiation transfer simulations for a toy explosion model where 
different viewing angles cause the late-time emission-line shift are presented, 
predicting a strong correlation between the color and shift, 
and a weaker one for the luminosity residual. 
\end{abstract}


\section{Introduction}
Type~Ia supernovae (SNe Ia) are  
used to measure cosmological parameters and study the nature of the dark energy 
(see Leibundgut\ 2008, and references therein). 
Thanks to the uniformity of their 
peak luminosities, once a phenomenological relation between 
the light-curve shape and the peak luminosity is applied 
(hereafter the light-curve correction, or the Phillips relation), 
they can be accurately used as cosmological standard candles 
(Phillips\ 1993; Hamuy et al.\ 1996; Phillips et al.\ 1999). 
Their colors are also known to correlate with light-curve shape 
(Tripp 1998; Tripp \& Branch 1999; Phillips et al.\ 1999).
In addition to light-curve shape and color, several other observables, mostly related to 
spectral features, have been shown to correlate with the SN peak luminosity 
(e.g., Nugent et al.\ 1995; Mazzali et al.\ 1998; Benetti et al.\ 2005; 
Bongard et al.\ 2006; Hachinger et al.\ 2006; Foley et al.\ 2008). 

Currently there are several issues yet to be clarified. One central issue is that 
the intrinsic color variations of SNe~Ia have not yet been fully understood. 
After application of the relation between color and  light curve shape, there remain 
variations in the color excess of SNe Ia. So far, it has been practically 
impossible to discriminate between the contributions from 
a possible `residual' intrinsic color, which does not correlate 
with the light curve shape, and that from the extinction within the host or the environment 
around the SN. 
This issue could be related to the fact that when the 
dispersion of the Hubble diagram is minimized with $R_{V}$ 
being treated as a free parameter, one obtains low values 
of $R_{V}$ between $1 - 2$. However, as shown by Folatelli 
et al. (2010; hereafter F10), when one compares the colors 
or color excesses of normal SNe~Ia, a more typical Milky Way-like 
value of the reddening law is obtained, i.e. $R_{V} \sim 3$.
F10 argued that this apparent discrepancy suggests that 
there is an intrinsic color variation within SNe~Ia that 
correlates with luminosity, but is independent of the light curve 
decline-rate $\Delta m_{15}$ (B).\footnote[1]{$\Delta m_{15} (B)$ is 
the magnitude difference in $B$-band 
between maximum brightness and 15 days later. } 
In the present study, we adopt an  $R_{V}$ $= 1.72$ as derived 
from F10 (i.e. Calibration 7 of Table~9).

Moreover, after application of the light-curve correction, Hubble diagram residuals at the level of 
$\sim 0.15$ mag exist for individual SNe~Ia
(e.g., Phillips et al.\ 1999; Prieto et al.\ 2006; 
Jha et al.\ 2007; Hicken et al.\ 2009ab). This is one of the 
issues that presently limit the 
precision in using SNe~Ia to constrain the value of the 
equation-of-state parameter of the dark energy (e.g., Hicken et al.\ 2009b; 
see also 
Wood-Vasey et al. 2007, Kessler et al. 2009 for the current status of the 
precision in SN~Ia cosmology). 
Several suggestions have been made for a 
secondary parameter that may provide a 
more accurate luminosity calibration\footnote[2]{Indeed, this is the `third' parameter, 
since light curve fitting methods usually use two parameters, i.e., the light curve shape 
and color. In this paper, we simply call the additional parameter the `second' parameter 
following the convention.} 
(or on parameters already including the effect of the second parameter). 
Suggestions include 
(i) metallicity (Gallagher et al.\ 2005; Timmes et al. 2003; Mazzali \& Podsiadlowski 2006; 
H\"oflich et al. 2010, but see also Howell et al.\ 2009; Neill et al.\ 2009; Yasuda \& Fukugita\ 2010), 
(ii) high-velocity spectral features (Wang et al.\ 2009b),  
(iii) spectral flux ratios (Bailey et al.\ 2009; Yu et al.\ 2009), 
and (iv) the mass and/or the morphological type 
of the host galaxy (Kelly et al.\ 2010; Lampeitl et al. 2010; Sullivan et al. 2010). 

An interesting possibility for the origin of the diverse properties of SNe Ia 
was recently suggested by Kasen et al. (2009) theoretically and by Maeda et al. 
(2010ab, hereafter M10a and M10b) observationally, namely  
an asymmetry in the SN explosion combined with the observer viewing angle.
In particular, M10a identified potential signatures 
of asymmetry in a number of SNe Ia, based on the observed wavelength 
shift of late-time emission lines (see \S 2 for more details). 
M10ab showed that the required configuration is qualitatively consistent with 
the expectation from a deflagration-to-detonation transition 
scenario\footnote[3]{In the deflagration-to-detonation transition scenario, the thermonuclear 
sparks first trigger the deflagration flames which travel subsonically, and then 
the flames subsequently turn into a supersonic detonation flame (Khokhlov 1991). } 
if the first thermonuclear 
sparks are ignited offset from the centre of the progenitor white dwarf. 

M10b suggested 
that the viewing angle effect is a probable origin of the spectral evolution diversity of 
SNe Ia. Different SNe~Ia show different velocity gradients ($\dot v_{\rm Si}$), 
defined as the speed of the decrease in the Si II absorption velocity 
after maximum brightness (Benetti et al. 2005; see also Branch et al. 1988). 
SNe are divided into high-velocity-gradient 
(HVG) ($\dot v_{\rm Si} > 70$ km s$^{-1}$) 
and low-velocity-gradient (LVG) objects 
($\dot v_{\rm Si} < 70$ km s$^{-1}$). 
M10b argued that different velocity gradients are a consequence 
of different viewing directions from which the SN is observed. It has 
been indicated 
that LVG and HVG SNe may show different properties in their intrinsic colors 
(e.g., Pignata et al.\ 2008) and that their luminosities may have to be calibrated 
in a different manner (Wang et al. 2009b). Here we revisit this color
issue in the context of our new interpretation of LVG and 
HVG SNe.

In this paper, we explore whether the late-time emission line shift, 
and thereby the observer viewing angle on an asymmetric explosion,
is related to the 
intrinsic color and the luminosity residuals of SNe~Ia after application of the Phillips relation. 
We find a correlation between the color at maximum brightness and the nebular emission 
line shift. 
We also investigate a possible relation between the luminosity residuals 
and the nebular line shifts, but since our sample is small the significance is not overwhelming. 
We then investigate the ramification of the viewing angle on the luminosity and color calibrations 
with the help of multi-dimensional radiation transfer calculations, 
and find that the predicted effect is qualitatively 
consistent with the trends seen in the data. 

The paper is organized as follows. 
In \S 2, we summarize the findings of M10a regarding the 
asymmetry in SNe~Ia, which are then used throughout the present paper.  
In \S 3, we present details of the sample of nearby SNe~Ia considered in this study. 
In \S 4, we discuss how the viewing angle is related to the intrinsic color of SNe~Ia. 
In \S 5, we discuss the procedures to estimate the intrinsic absolute magnitude and 
subsequent residuals. 
In \S 6, we compare the late-time emission line shifts with the luminosity residuals. 
In \S 7, we investigate the effect of the viewing angle on the peak brightness and color 
by simulating light curves for kinematic off-centre toy models. 
In \S 8 the paper is closed with conclusions, discussion and future perspectives. 

\section{Asymmetry in Type Ia SNe}

\begin{figure*}
   \centering
   \includegraphics[width=0.45\textwidth]{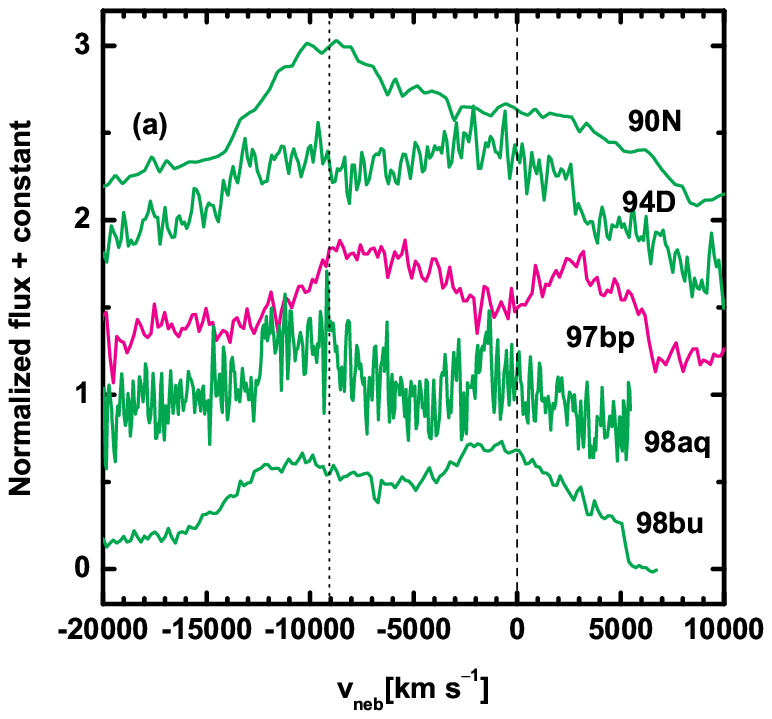}
   \includegraphics[width=0.45\textwidth]{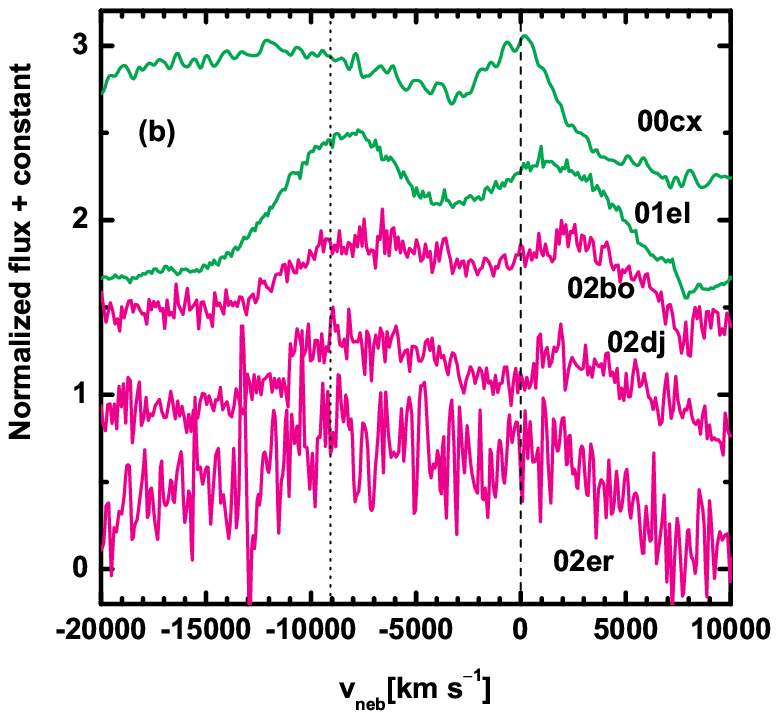}\\
   \includegraphics[width=0.45\textwidth]{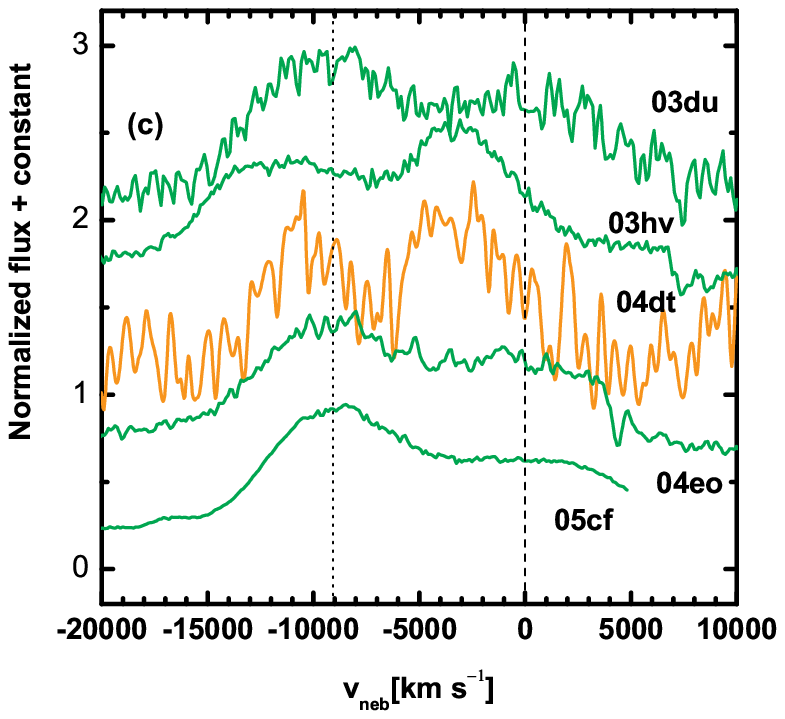}
   \includegraphics[width=0.45\textwidth]{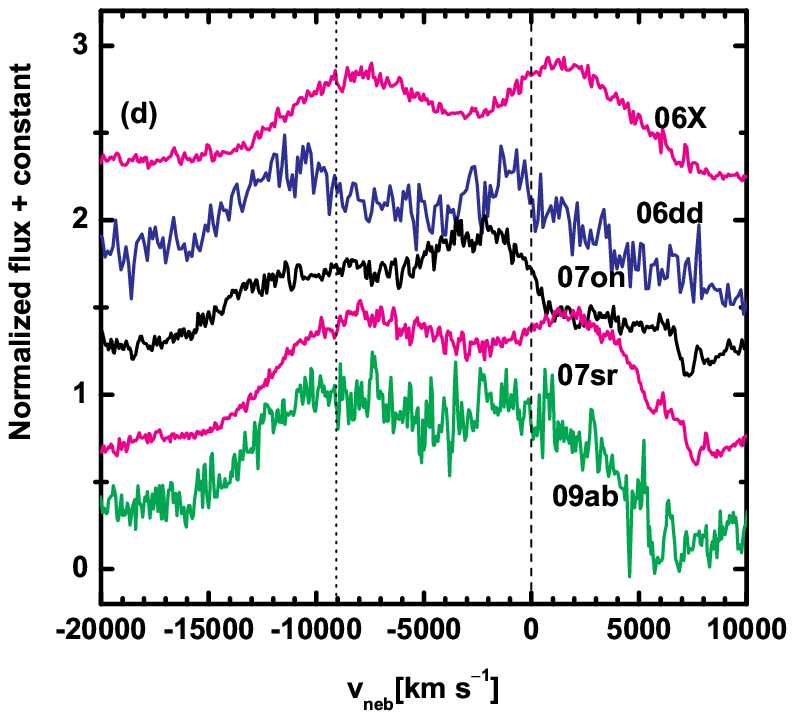}
   \caption{[Fe II]~$\lambda$7155 and [Ni II]~$\lambda$7378 
in late-time spectra of SNe Ia (see references in Table 1). 
The spectra have been redshift-corrected, and then converted to 
a velocity assuming 7378~\AA\ as the zero-velocity. The rest positions of [Fe II] and [Ni II] 
are denoted by dotted and dashed lines, respectively. The color coding indicates the 
early-phase `velocity gradient' (see \S 1; Benetti et al. 2005): 
Low velocity gradient SNe (LVG SNe) 
in green, high velocity gradient SNe (HVG SNe) in magenta. SN 2007on is 
a fast declining SN 1986G-like SN Ia (Morrell et al. 2007) and is shown in black. 
SN 2006dd, without a measured velocity gradient, is shown in blue. SN 2004dt, 
which was argued to be a peculiar outlier based on late-time spectra (M10b), is shown in orange.} 
   \label{fig1}
\end{figure*}

\begin{figure*}
   \centering
   \includegraphics[width=0.45\textwidth]{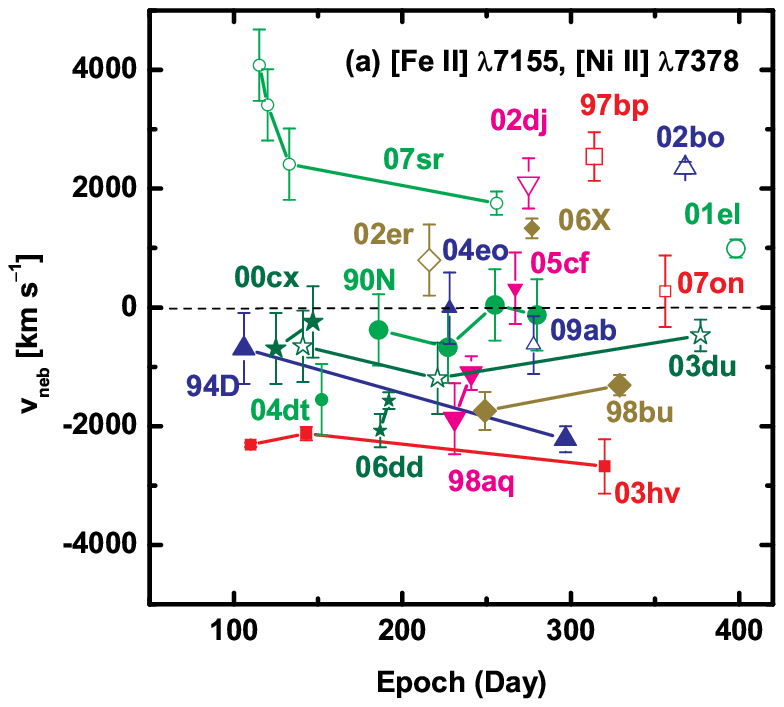}
   \includegraphics[width=0.45\textwidth]{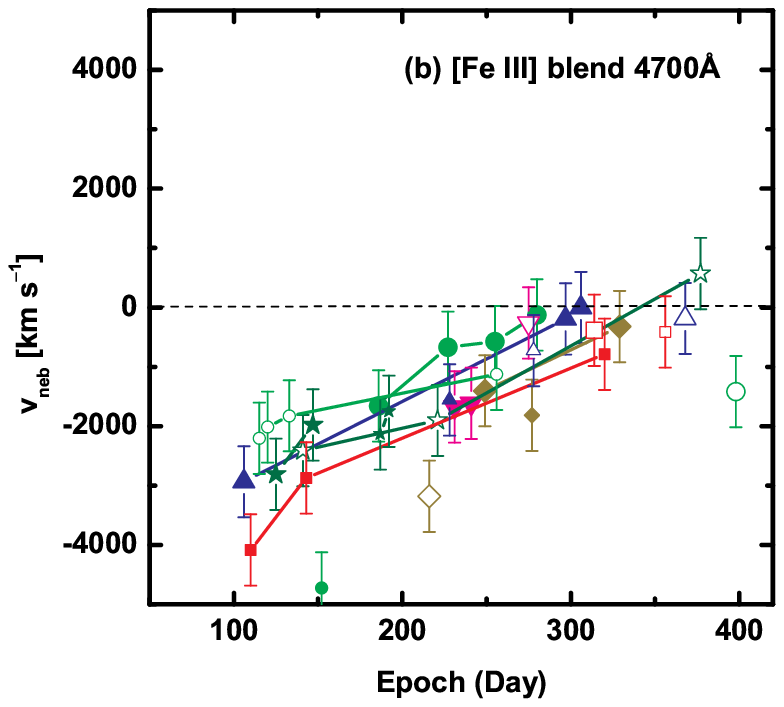}
   \caption{Observed wavelength shifts for (a) the 
[Fe II]~$\lambda$7155 and [Ni II]~$\lambda$7378 lines, and (b) the [Fe III] blend at $4700$~\AA, 
plotted as a function of time since $B$-band maximum. 
The wavelengths have been converted to velocities, assuming the expected rest wavelengths of the lines as the zero-velocity.  The same objects are connected by  lines. 
The data are from our present sample (see \S 3, Table 1 and references therein). 
}
   \label{fig2}
\end{figure*}

In this section, we summarize the findings presented in M10a regarding 
asymmetry in SNe Ia. 
M10a suggested that the innermost region of the ejecta, filled with stable Fe-peak elements, 
is generally offset, and that an observer's viewing angle can be traced by 
shifts in the central wavelengths of nebular emission lines of [Fe II]~$\lambda$7155 and 
[Ni II]~$\lambda$7378 at late phases. The argument is as follows: 
\begin{enumerate}
\item Using a sample of late-time (i.e., at least 100 days after maximum brightness) 
SN~Ia spectra, mostly drawn from the SUSPECT 
database,\footnote[4]{The Online Supernova Spectrum Archive, SUSPECT, is found 
at http://bruford.nhn.ou.edu/$^\sim$suspect/ .}
M10a found that in cases where the  [Fe II]~$\lambda7155$ and [Ni II]~$\lambda$7378 lines were detected, they exhibited measurable shifts in their central wavelength with respect to 
the expected rest wavelength, as shown in Fig. 1.  
There is no clear correlation between the phase of the nebular spectrum and the shift of 
the [Fe II]~$\lambda$7155 and [Ni II]~$\lambda$7378 lines. In addition, no significant 
evolution is seen in the measured line shifts for individual SNe~Ia at $\gsim 150$ days. 
To illustrate these characteristics, we show the wavelength shifts 
(converted to a velocity) of these lines in Fig.~2. 
\item On the other hand, the strongest lines at late phases, i.e., the [Fe III] blend at 
$\sim 4700$~\AA\ and the 
[Fe II]/[Fe III] blend at $\sim 5250$~\AA, behave similarly for all SNe~Ia. 
Since the [Fe III] blend at $\sim 4700$~\AA\ is stronger and its blended 
nature can be better handled (M10a), we show its temporal evolution in Fig.~2.
There is a clear correlation between the phase the nebular spectrum 
and the central wavelength of the [Fe III] 
blend at 4700~\AA, unlike for [Fe II]~$\lambda$7155 and [Ni II]~$\lambda$7378.  
The shifts of [Fe III] at $4700$~\AA\ evolve from the blue to the rest 
wavelength for all SNe. 
\item The temporal behavior seen in Figure 2 suggests 
that the shifts of [Fe II]~$\lambda$7155 
and [Ni II]~$\lambda7378$ trace the line-of-sight velocity of the emitting material.  
As a consequence, the diversity in the wavelength shifts indicates that the distribution of the emitting material is offset with respect to the centre of the SN ejecta. 
Note that spherically symmetric explosions with different expansion 
velocities should produce different line widths, but never a shift in the central wavelength. 
As a result of the offset, the observed diversity  arises from different observer viewing angles. 
On the contrary, the observed features of 
the [Fe III] blend at 4700~\AA\ 
suggest that the [Fe III] blend traces the 
distribution of the emitting material only at sufficiently late epochs, and 
the small diversity indicates that the emitting material is distributed more or less spherically. 
\end{enumerate}

M10a argued that these characteristics can be naturally explained by 
a deflagration-to-detonation explosion scenario, 
if the distribution of the inner deflagration ash has an offset and the outer 
detonation ash is distributed in a roughly spherically symmetric way. 
They proposed that these lines are emitted from 
different regions -- an outer, relatively low-density region dominated by radioactive 
$^{56}$Ni (which decays into $^{56}$Co and then into $^{56}$Fe, powering the SN light curve), 
and an inner, relatively high-density region dominated by stable Fe-peak elements, 
i.e., $^{58}$Ni, $^{56}$Fe, and $^{54}$Fe (see also Mazzali et al. 2007). 
The former and latter correspond to the detonation and the deflagration ashes, 
respectively. 
For such a configuration, 
the outer region should be in a relatively high ionization state (i.e., doubly-ionized) and 
at high temperature (electron temperature $T_{\rm e} \gsim 10,000$~K). 
On the other hand, the inner region should be in a low ionization state (i.e., singly-ionized) 
and at low temperature ($T_{\rm e} \sim 2000 - 7000$~K). This stems from 
(a) the ionization balance and (b) the thermal balance. 
Namely, 
(a) $n_{i+1}/n_{i} \propto J_{\gamma} n_{\rm e}^{-1}$, where 
$n_{i}$ is the density of the $i$-th ionization state, $J_{\gamma}$ is the 
radioactive energy input from the decay chain $^{56}$Ni $\to$ $^{56}$Co $\to$ 
$^{56}$Fe, and $n_{\rm e}$ is the electron density. (b) 
$e^{-T_{\rm ex}/T_{\rm e}} \propto J_{\gamma} n_{\rm e}^{-1} n_{0}^{-1}$, 
where $T_{\rm ex}$ is the excitation temperature 
of the line, and $n_{0}$ is the population of the lower level. 

The [Fe III] blend at $4700$~\AA\ is emitted by Fe$^{++}$ with high $T_{\rm ex}$, 
and thus the outer region dominates the emission of this blend. 
This blend is thus attributed to emission from the detonation ash in the 
deflagration-to-detonation transition scenario. 
The opposite is true for 
[Fe II]~$\lambda$7155 and [Ni II]~$\lambda$7378, 
and thus these lines are mostly emitted from the inner region, 
i.e., the deflagration ash. 
Combined with the phenomenologically derived distribution of material 
emitting these lines (see above), 
this argues that the deflagration ash is located offset from the centre of the progenitor star, 
while the detonation ash is distributed spherically. 

Although these arguments are based on a few lines in the optical wavelength regime, 
this interpretation predicts that, at late phases, the lines from high ionization states 
and/or with high excitation temperature should show virtually no shift, while those from low ionization 
states and/or with low excitation temperature show a wavelength shift depending on the viewing 
direction. M10a showed that this is 
the case for SN 2003hv, which has 
a well observed spectrum at late phases, 
covering spectral lines from the optical all the way to the mid-infrared (IR)
(Motohara et al.\ 2006; Gerardy et al.\ 2007; Leloudas et al.\ 2009). 
The size of the offset derived for SN 2003hv was found to be $\sim 3,500$ km~s$^{-1}$ (M10a). 
Note that this behavior in various lines from optical through mid-IR provides 
an additional argument against line blending and radiation transfer as a cause of the 
observed line shifts. 

\begin{table*}
\centering
 \begin{minipage}{160mm}
 \caption{SN Ia sample}
 \label{tab:sample}
 \scriptsize
 \begin{tabular}{@{}llllllclrl@{}}
 \hline
 SN & 
Host & 
z & 
$\mu$\footnote{Distance modulus from 
KP Cepheid measurements (KP), Surface Brightness Fluctuations 
(SBF), the near-IR Tully-Fisher relation (TF), 
the planetary nebulae luminosity function (PNLF), 
the tip of the giant branch (TRGB), 
or the host-galaxy recession velocity corrected 
for the Virgo infall (HF). See \S 5 for details and references.} & 
$\Delta m_{15}(B)$\footnote{Obtained from the literature, except for 
SNe 2007sr, 2007on, and 2009ab (Stritzinger et al.\ in prep.). 
For these 3 SNe, $\Delta m_{15} (B)$ estimates were obtained 
using the multi-color template light curve fitter SNooPy (Burns et al.\ 2011). 
} & 
$V$\footnote{The values at maximum brightness. Corrected for Galactic extinction.} & 
$B-V$$^{\rm c}$& 
References\footnote{References for the photometric properties. 
A05. Anupama et al. (2005); 
A07. Altavilla et al. (2007); 
B03. Branch et al. (2003); 
B04. Benetti et al. (2004); 
C01. Cappellaro et al. (2001);
C03. Candia et al. (2003);  
G96. G\'omez et al. (1996); 
H00. Hernandez et al. (2000); 
J99. Jha et al. (1999); 
K03. Krisciunas et al. (2003); 
K05. Kotak et al. (2005); 
L98. Lira et al. (1998); 
L01. Li et al. (2001); 
L07. Leonard (2007); 
L09. Leloudas et al. (2009); 
M05. Mattila et al. (2005); 
M10b. Maeda et al. (2010b); 
P96. Patat et al. (1996); 
P04. Pignata et al. (2004); 
P07a. Pastorello et al. (2007a); 
P07b. Pastorello et al. (2007b); 
P08. Pignata et al. (2008); 
R95. Richmond et al. (1995); 
R99. Riess et al. (1999); 
R05. Reindl et al. (2005); 
S04. Sollerman et al. (2004); 
S05. Stehle et al. (2005); 
S07. Stanishev et al. (2007);  
S10a. Stritzinger et al. (2010); 
S10b. Stritzinger et al. (in prep.); 
W08. Wang et al. (2008); 
W09. Wang et al. (2009a); 
Y09. Yamanaka et al. (2009)
} & 
$v_{\rm neb}$ & 
References\footnote{References for the late-time spectra. 
}\\
   &  &   &   (mag)  &  (mag)  & (mag) & (mag) &   & (km s$^{-1}$) &  \\ 
\hline
1990N & NGC 4639 & 0.003395 & $31.71 \pm 0.15$ (KP) & $1.07 \pm 0.05$ & 
12.62 & 0.037 & L98 & $-126 \pm 600$ & G96\\
1994D & NGC 4526 & 0.001494 & $30.98 \pm 0.20$ (SBF) & $1.32 \pm 0.05$ & 
11.83 & -0.080 & R95, P96 & $-2220 \pm 220$ & G96\\
1997bp & NGC 4680 & 0.008312 & $32.68 \pm 0.27$ (HF) & $0.97 \pm 0.2$ & 
13.78 & 0.16 & R05 & $2539 \pm 410$ & M10b\\
1998aq & NGC 3982 & 0.003699 & $31.56 \pm 0.08$ (KP) & $1.12 \pm 0.05$ & 
12.42 & -0.11 & R99 & $-1106 \pm 286$ & B03\\
1998bu & NGC 3368 & 0.002992 & $30.11 \pm 0.20$ (KP) & $1.06 \pm 0.05$ & 
11.80 & 0.34 & J99, H00 & $-1309 \pm 171$ & C01\\
2000cx & NGC 524 & 0.007935 & $32.63 \pm 0.27$ (HF) & $0.93 \pm 0.04$ & 
12.99 & 0.10 & L01& $-244 \pm 600$ & C03, S04\\
2001el & NGC 1448 & 0.003896 & $31.23 \pm 0.45$ (TF) & $1.13 \pm 0.04$ & 
12.69 & 0.068 & K03 & $993 \pm 152$ & M05\\
2002bo& NGC 3190 & 0.004240 & $31.70 \pm 0.24$ (SBF)\footnote{An SBF distance to NGC 3226, 
a member of the same group. } & $1.16 \pm 0.03$ & 
13.50 & 0.44 & B04 & $2350 \pm 100$ & S05\\
2002dj & NGC 5018 & 0.009393 & $32.82 \pm 0.25$ (HF) & $1.08 \pm 0.05$ & 
13.85 & 0.063 & P08 & $2090 \pm 423$ & P08\\
2002er & UGC 10743 & 0.008569 & $32.87 \pm 0.25$ (HF) & $1.32 \pm 0.03$ & 
14.10 & 0.16 & P04 & $797 \pm 600$ & K05\\
2003du & UGC 9391 & 0.006384 & $32.42 \pm 0.30$ (HF) & $1.02 \pm 0.05$ & 
13.54 & -0.089 & A05, S07 & $-471 \pm 265$ & S07\\
2003hv & NGC 1201 & 0.005624 & $31.37 \pm 0.30$ (SBF) & $1.61 \pm 0.02$ & 
12.49 & -0.035 & L09 & $-2677 \pm 457$ & L09\\
2004dt & NGC 799 & 0.019730 & $34.55 \pm 0.12$ (HF) & $1.21 \pm 0.05$ & 
15.25 & -0.053 & A07 & $-1551 \pm 600$ & A07\\
2004eo & NGC 6928 & 0.015701 & $34.12 \pm 0.14$ (HF) & $1.45 \pm 0.04$ & 
15.02 & 0.064 & P07a & $-14 \pm 600$ & P07a\\
2005cf& MCG-1-39-3 & 0.006461 & $32.19 \pm 0.33$ (HF) & $1.07 \pm 0.03$ & 
13.26 & -0.017 & P07b, W09 & $324 \pm 600$ & L07\\
2006X & NGC 4321 & 0.005240 & $30.91 \pm 0.20$ (KP) & $1.31 \pm 0.05$ & 
13.96 & 1.34 & W08, Y09 & $1331 \pm 164$ & W08\\
2006dd & NGC 1316 & 0.005871 & $31.26 \pm 0.10$ (PNLF) & $1.08 \pm 0.03$ & 
12.32 & -0.060 & S10a & $-1569 \pm 142$ & S10a\\
2007on & NGC 1404 & 0.006494 & $31.45 \pm 0.19$ (SBF) & $1.55 \pm 0.01$ & 
12.98 & 0.12 & S10b & $272 \pm 600$ & S10b\\
2007sr & NGC 4038 & 0.005477 & $31.51 \pm 0.17$ (TRGB) & $1.12 \pm 0.01$ & 
12.54 & 0.12 & S10b & $1754 \pm 198$ & S10b\\
2009ab & UGC 2998 & 0.011171 & $33.29 \pm 0.20$ (HF) & $1.25 \pm 0.02$ & 
14.60 & 0.058 & S10b & $-634 \pm 486$ & S10b\\
\hline
\end{tabular}
\end{minipage}
\end{table*}

This finding was interpreted as evidence that the initial deflagration 
flame proceeds in an asymmetric way, having an offset with respect 
to the explosion centre, in the context of a deflagration-to-detonation 
transition explosion scenario (Khokhlov\ 1991; Yamaoka et al.\ 1992; 
Woosley \& Weaver\ 1994; Iwamoto et al.\ 1999; R\"opke \& Niemeyer\ 2007; 
Kasen et al.\ 2009; Seitenzahl et al.\ 2010). 
Although the details of the ignition process are not yet fully understood, theoretically 
an off-centre ignition may be a natural consequence of convection within 
a progenitor white dwarf (Kuhlen et al. 2006). 
Maeda et al. (2010c) (hereafter M10c) 
argued, based on their hydrodynamic and nucleosynthesis simulations, 
that a deflagration-to-detonation transition model with the initial sparks ignited 
at an offset from centre will result in a configuration qualitatively 
consistent with the above findings. 

According to this interpretation and an extensive search of emission lines 
from the optical to mid-IR wavelengths, 
M10a suggested that the best lines to use in the optical 
for probing the asymmetry in the innermost SN ejecta are [Fe II]~$\lambda$7155 
and [Ni II]~$\lambda$7378 
(as well as [Fe II]~$\lambda$8621; Leloudas et al.\ 2009).  
These are the only ones which satisfy the criteria that (a) they reflect 
the low ionization and 
low temperature, (b) they suffer from little blending, and (c) they are covered 
by most optical spectra. We therefore adopt these lines as diagnostics of the 
ejecta asymmetry and the viewing angle in the present study.

\section{Supernova Sample}

In this paper, we investigate how an asymmetric explosion and different viewing angles 
influence the color and luminosity of SNe~Ia, by examining 
correlations between these quantities and 
the emission line shift in the late-time spectra. The sample of SNe~Ia is thus limited 
by the requirement that late-time nebular spectra are available. 
Our initial late-time spectral data set comprised the same 20 SNe~Ia used by M10a, 
mostly drawn from the SUSPECT database,
supplemented by other late-time nebular spectra accessible to the authors.
To include these spectra in the present analysis, we imposed some additional criteria: 
(i) the spectrum was taken at $\gsim 150$ days after maximum brightness (see Fig. 2), 
(ii) either [Fe II]~$\lambda$7155 or [Ni II]~$\lambda$7378 could be identified, 
(iii) the spectra had to be of high enough quality that the central wavelength 
of either of these lines could be measured, 
(iv) the SNe were required  to have good early phase $B$- and $V$-band photometry so 
that  their light-curve parameters and color at maximum could be estimated,
and 
(v) $\Delta m_{15} (B)$ was in the range between $0.7$ to $1.7$ mag 
in order to apply the well-developed relations between $\Delta m_{15} (B)$ and 
the peak absolute magnitude/color (e.g., Phillips et al. 1999; F10). 

From the initial data set, 13 objects fulfilled these criteria. 
In addition, we included 7 SNe 
which satisfied our criteria, i.e. SNe~1997bp, 2002bo, 2006X, 
2006dd, 2007on, 2007sr, and 2009ab. 
In 13 out of these 20 objects, both [Ni II]~$\lambda$7378 and [Fe II]~$\lambda$7155 
were securely identified (see below). In the other 7 events, only one of these lines 
was discernable. 
These 7 SNe~Ia were included in our analysis because a measurement 
of their nebular line velocity shift was possible, 
although less secure than when both lines were identified. 
The sample of 20 SNe~Ia used in the present study is listed in Table~1. 
Their late-time spectra in the wavelength range covering the [Fe II] and [Ni II] 
are shown in Figure 1. 
Details on SN 2006dd are presented in Stritzinger et al. (2010), while 
SNe 2007on, 2007sr, and 2009ab 
will be presented elsewhere. 

M10a measured the wavelength shifts 
in [Ni II]~$\lambda7378$ at late phases 
for the majority of the SNe~Ia listed in Table~1. 
In M10b, they also measured the shift in [Fe II]~$\lambda$7155.  
In this study, we remeasured these line shifts and 
updated the velocity shift ($v_{\rm neb}$), as follows: 
when both [Ni II]~$\lambda7378$ and [Fe II]~$\lambda$7155 
were identified, we fitted their central wavelengths with Gaussian profiles 
simultaneously using the {\tt IRAF} {\em deblend} 
command.\footnote[5]{IRAF is distributed by 
the National Optical Astronomy Observatories, 
which are operated by the Association of Universities for Research 
in Astronomy, Inc., under cooperative agreement with the National 
Science Foundation.} 
The $v_{\rm neb}$ is taken to be the mean value of them 
(see also M10b).\footnote[6]{The two lines were not simultaneously fitted in the 
previous measurements, and this changed the exact values of 
$v_{\rm neb}$. However, the old and new measurements 
are mostly consistent within the associated errors. 
The only exception is SN 1990N, for which the host galaxy redshift 
was incorrectly treated in the previous measurement. We confirmed that the conclusions of M10ab are not affected by the 
difference in the measurement in $v_{\rm neb}$. } 
As a default estimate, the continuum is taken to be constant across 
the wavelength range. We estimated the error in the fit by 
varying the continuum. We further considered the 
errors arising from the difference in the velocity shifts of 
the two lines. The regions emitting these two lines 
are not exactly the same since the region emitting [Ni II]~$\lambda$7378 is attributed to be 
a product of the deflagration at the very beginning, while that emitting 
[Fe II]~$\lambda$7155 is a product from a later phase, but still before the ignition of the detonation 
(M10a; see also M10c for an example of a detailed nucleosynthesis study). 
Therefore, there could be intrinsic differences 
in the velocity shifts of these two lines. The final error bars are taken to be the larger 
one between the errors associated with the Gaussian fit and the difference in the two lines. 

In the majority of SNe,  
The shifts seen in [Ni II]~$\lambda$7378 agree with those measured 
in [Fe II]~$\lambda$7155  to within an error of at most $\sim 600$ km s$^{-1}$. 
If only one of these lines could be identified, we estimated the velocity shift only 
from the single line using the (single-line) Gaussian fitting command within {\tt IRAF}, 
and assumed that the error is $600$ km s$^{-1}$. 
In the case of the fast-declining 1986G-like SN~2007on (Morrell et al. 2007) 
and the peculiar (in terms of 
the late-time spectra) SN 2004dt (M10b), we did not use the [Ni II] feature
as it might be contaminated or originate from [Ca II]~$\lambda\lambda$7291, 7324  
(Filippenko et al. 1992; Turatto et al. 1996). 
Indeed, we suspected that [Ca II]~$\lambda\lambda$7291, 7324 
might contribute much to the ``[Ni II] feature'' 
for SNe 1994D, 2003hv, 2004dt, and 2007on, 
since the measured position of the [Ni II] feature is close to the rest wavelength 
of the [Ca II]. 
Among these SNe,  we regard that the [Ni II] identification is the case for SNe 1994D and 2003hv, 
since its velocity shift agrees with that measured from the [Fe II]. On the other hand, 
these velocities are inconsistent for SNe 2004dt and 2007on, thus we believe 
that [Ca II] is a more likely identification for these two SNe. 
This indicates that the strong [Ca II] in the late phase may be a property shared by 
a part of fast declining SNe and related objects, but not by Branch-normal SNe.  
When spectra at several epochs were available, we adopted $v_{\rm neb}$ 
estimated from the latest epoch for which the quality of the spectrum 
was sufficiently high, since a more secure estimate of the intrinsic velocity shift 
for the nebular emission lines is possible using later phase data (\S 2). 
As shown in Fig.~2, $v_{\rm neb}$ measured for the same 
SN at different epochs agrees within the errors, at least for $\gsim 150$ days.

\section{Intrinsic Color Variations}

\begin{figure}
   \centering
   \includegraphics[width=0.45\textwidth]{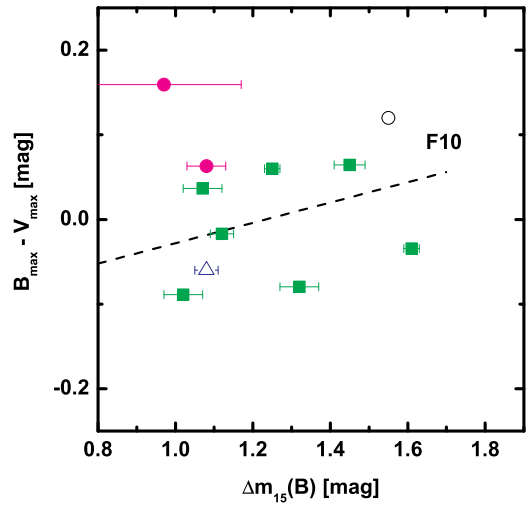}
   \caption{The $B_{\rm max}-V_{\rm max}$ color of 11 ``low-extinction'' SNe as a function of 
$\Delta m_{15} (B)$. The color is corrected for Galactic extinction only. 
SNe are shown by different symbols 
defined as follows: Low velocity gradient SNe (LVG SNe) by green filled-squares, 
high velocity gradient SNe (HVG SNe) by magenta filled-circles, 
the fast-declining SN 2007on by a black open-circle. 
SN 2006dd, for which no information on its velocity gradient is available, 
is shown by a blue open-triangle. The 
relation between color and 
$\Delta m_{15} (B)$ as derived by F10 is shown 
by a dashed line. 
}
   \label{fig3}
\end{figure}

\begin{figure*}
   \centering
   \includegraphics[width=0.45\textwidth]{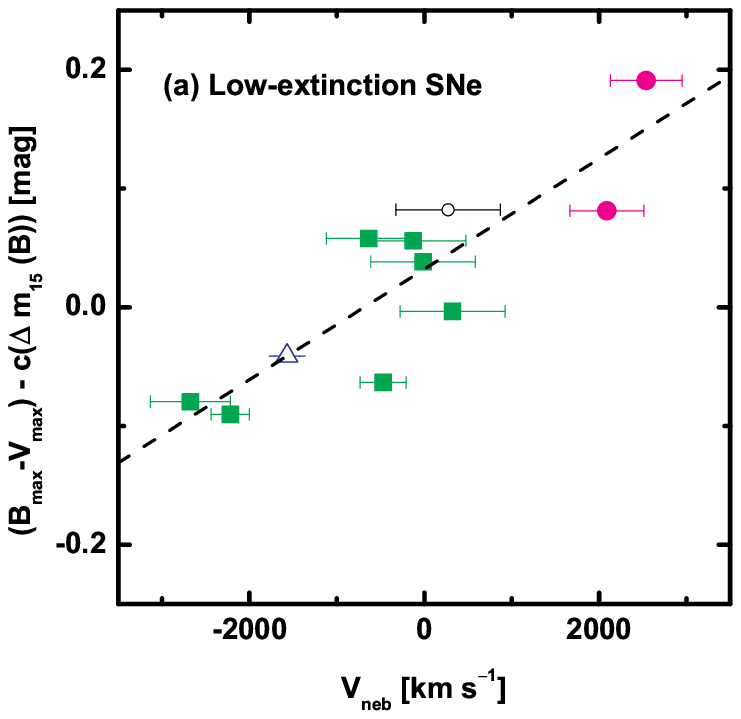}
   \includegraphics[width=0.45\textwidth]{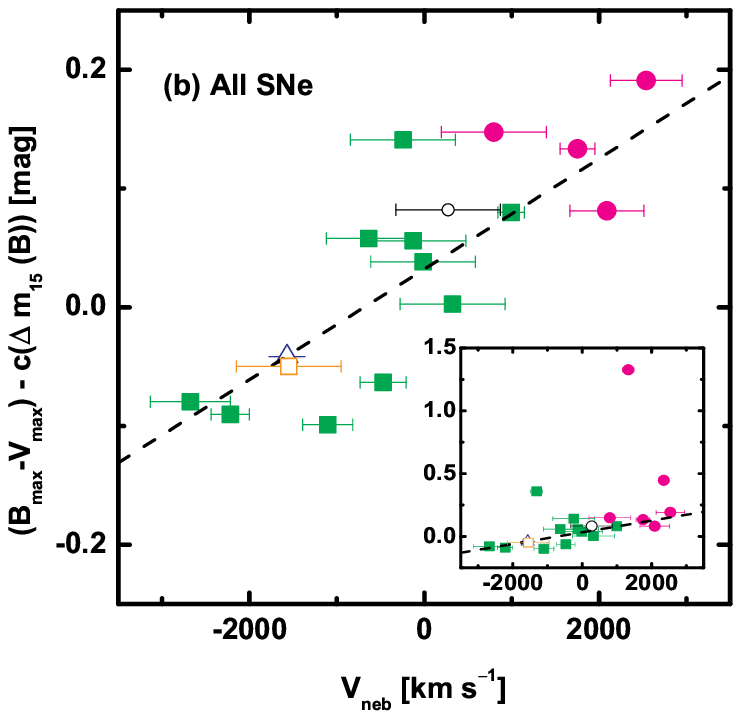}
   \caption{The $B_{\rm max}-V_{\rm max}$ color, {\it after}  correction of the 
$\Delta m_{15} (B)$ - color relation of F10 
($c(\Delta m_{15} (B))$: see the main text), 
vs. $v_{\rm neb}$. 
The dashed line is the best linear fit to the 11 low-extinction SNe.  
(a) shows the low-extinction sample, and (b) the entire sample 
(zoomed in the same region as in the panel (a)). 
The inset in (b) shows the entire sample, including also 
the 3 highly-reddened SNe Ia.
SNe are shown by different symbols 
defined as follows: Low velocity gradient SNe (LVG SNe) by green filled-squares, 
high velocity gradient SNe (HVG SNe) by magenta filled-circles, and 
the fast-declining SN 2007on by a black open-circle. 
SN 2004dt, which was suggested to be a peculiar outlier based on its late-time 
spectrum, is shown by an orange open-square. 
SN 2006dd, for which no information on its velocity gradient is available, 
is shown by a blue open-triangle. 
}
   \label{fig4}
\end{figure*}

We first examine whether the intrinsic colors of SNe~Ia are related 
to $v_{\rm neb}$ and therefore to the viewing direction of an observer 
in the asymmetric explosion scenario. 
For this purpose, we define the {\em pseudo color\,} ($B_{\rm max} - V_{\rm max}$) as 
the difference in peak $B$ and $V$ magnitudes obtained at the time of 
maximum light {\em in each bandpass}. 

We first identify a ``low-extinction'' sample of SNe 
for which the extinction within the host galaxy is likely insignificant. We regard SNe as 
low-extinguished if (i) the host is an elliptical or S0 galaxy or (ii) the SN was located in 
the outskirts, further from the centre 
than half of the apparent radius of the host (defined as the length of the 
projected major axis at the isophotal level 25 mag arcsec$^{-2}$; Paturel et al. 1991)
as obtained from the Lyon-Meudon Extragalactic Database (Paturel et al.\ 2003).  
We also remove SN 2000cx, since this was reported to show peculiar colors (Li et al. 2001; 
Candia et al. 2003).
Our adopted ``low-extinction'' sample then comprises 11 objects. 

\begin{figure*}
   \centering
   \includegraphics[width=0.45\textwidth]{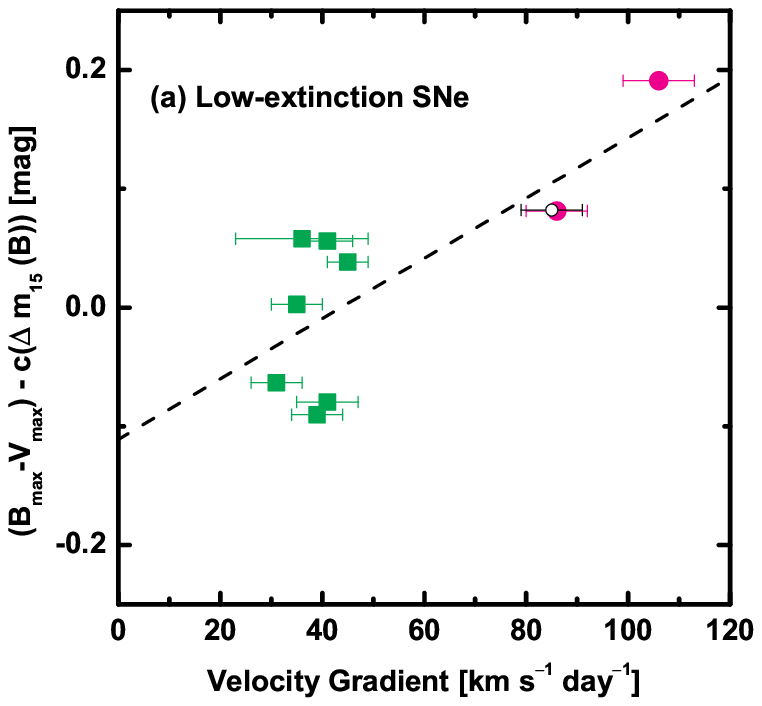}
  \includegraphics[width=0.45\textwidth]{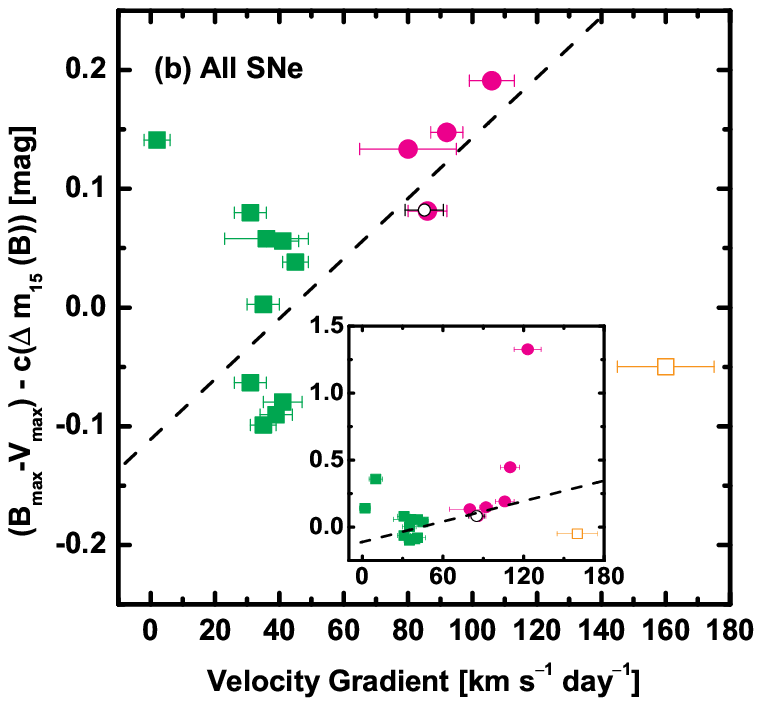}
   \caption{The $B_{\rm max}-V_{\rm max}$ color, {\it after} the correction of the 
$\Delta m_{15} (B)$ - color relation of F10 
($c(\Delta m_{15} (B))$; see the main text), 
as a function of the velocity gradient $\dot v_{\rm Si}$. 
The dashed line is the best linear fit to the 10 low-extinction SNe 
(excluding SN 2006dd for which the velocity gradient is not available).  
The symbols and color coding indicate 
the velocity gradient near maximum brightness (see Fig. 4 caption). 
}
   \label{fig5}
\end{figure*}

It has been shown that the 
colors of SNe~Ia at maximum brightness are related to $\Delta m_{15} (B)$ such 
that SNe with a larger decline rate are intrinsically redder (e.g., Phillips et al. 1999). 
Figure 3 shows a comparison between 
$\Delta m_{15} (B)$ and the pseudo color 
($B_{\rm max}-V_{\rm max}$) for our low-reddening sample, 
corrected for Galactic extinction adopting the values derived from the dust maps of 
Schlegel et al. (1998) and $R_{V} = 3.1$. 
Following the suggestion by M10b that the viewing direction is related to the velocity gradient, 
we hereafter plot the SNe with different symbols depending on whether 
they belong to the LVG or the HVG groups.
Although we do not see a strong correlation, 
our data are overall consistent with the $\Delta m_{15} (B)$ - color relation of the 
``low-reddened'' sample of F10 defined as 
$c (\Delta m_{15} (B)) \equiv 
(B_{\rm max} - V_{\rm max})_0 = -0.016 + 0.12 [\Delta m_{15} (B) - 1.1]$ mag.

Figure~4 shows a comparison between $v_{\rm neb}$ and pseudo color corrected 
for Galactic extinction {\em and} the color vs. $\Delta m_{15} (B)$ relation of F10. 
In other words, we are investigating a relation between $v_{\rm neb}$ and the `color 
residual' after correcting for the $\Delta m_{15} (B)$ term. 
Figure~4 reveals 
that there is a clear correlation between $v_{\rm neb}$ and this color 
for the low-extinction sample. 
A linear fit to the 11 SNe results in 
a chance probability of only $4.4 \times 10^{-4}$ (i.e., $3.3\sigma$ significance). 
The standard deviation is $0.05$ mag. 
Figure 4 further shows that almost the same relation applies even when not 
restricted to the low-extinction SNe (but omitting 3 evidently highly-reddened SNe). This suggests that 
this $v_{\rm neb}$ - color relation is at least as important as the $\Delta m_{15} (B)$ 
correction in the color. It further indicates that a large part of the color variations 
of SNe Ia can be attributed to intrinsic variations due to different viewing directions, 
rather than the host extinction. 
The intrinsic color we  derive can be expressed by the following equation: 
\begin{eqnarray}
(B_{\rm max} - V_{\rm max})_0 
& =  & 0.016 + 0.12 [\Delta m_{15} (B) - 1.1] \nonumber\\
& + & 0.047 (v_{\rm neb} / 1000 \ {\rm km} \ {\rm s}^{-1}) \ {\rm mag}.
\end{eqnarray}

It has been suggested that some HVG SNe may be intrinsically redder than 
LVG SNe (e.g., Pignata et al. 2008). 
Wang et al. (2009b) showed that SNe with higher Si II absorption velocity 
(which is a typical signature of HVG SNe) are intrinsically redder than those with lower velocity 
(roughly corresponding to LVG SNe) by $\sim 0.1$ mag. Therefore, we qualitatively 
expect that SNe with larger (positive) $v_{\rm neb}$ are intrinsically redder according to M10b, 
which is consistent with the data shown here. 
This also explains why Wang et al. (2009b) obtained a smaller dispersion 
in the calibrated luminosities by assuming a smaller $R_{V}$ for SNe with higher velocity 
in the Si II absorption (i.e., roughly equivalent to HVGs). If these SNe are intrinsically redder than 
the others, then a smaller $R_{V}$ mimics the effect of the viewing direction.

Figure 5 shows a comparison between the velocity gradient and $B_{\rm max} - V_{\rm max}$ 
(corrected for Galactic extinction and the $\Delta m_{15} (B)$ term). As suggested previously, 
HVG SNe are redder than LVG SNe. The linear fit results in $P = 0.0073$ ($2.4\sigma$) 
for the low-reddened SNe. 
We note that the distribution can actually be better expressed by a bimodal distribution 
(HVG SNe and LVG SNe) rather than a continuous distribution. 
Velocity gradients are similar for all LVG SNe, whereas we can see a color variation 
as a function of $v_{\rm neb}$ also within this group. 
The correlation with the velocity gradient is indeed weaker than the relation using $v_{\rm neb}$. 
This suggests that $v_{\rm neb}$ is a better indicator of the intrinsic color than the velocity gradient.
This mostly stems from the saturation of the velocity gradients 
for LVG SNe, which makes it difficult to derive the detailed 
viewing direction directly from the velocity gradient (M10b). 

Our finding that color is strongly correlated with $v_{\rm neb}$ 
suggests that the viewing direction is indeed an important property which controls the color as 
theoretically investigated by Kasen et al. (2009) (see also Foley \& Kasen 2010). 
M10b attributed the variation in velocity gradients in different 
radial density distributions at different directions; they suggested 
that the photospheric velocity is smaller and evolves more slowly for a direction 
closer to the offset ignition direction. 
A smaller photospheric velocity ($v_{\rm ph}$) leads to a larger photospheric temperature 
($T_{\rm eff}$) since $T_{\rm eff} \propto v_{\rm ph}^{-1/2}$. 
The larger $T_{\rm eff}$ results in a bluer color (c.f., figs. 1 and 3 of Nugent et al. 1995). 
Also, the difference in $T_{\rm eff}$ has been suggested to be a main origin of different spectral 
features seen in HVG and LVG SNe (Hachinger et al. 2008; Tanaka et al. 2008). 
This qualitatively explains the tendency we have identified in the data. 

This result opens up a possibility to estimate the host extinction 
by including the effect of viewing angle on different SNe:
we compute the color excess due to host extinction as $E (B-V) \equiv (B_{\rm max} - V_{\rm max}) - (B_{\rm max} - V_{\rm max})_0$, where
the SN intrinsic color 
$(B_{\rm max} - V_{\rm max})_0$ is expressed by Equation 1. 
Note that some SNe then show small negative color excesses. since Equation 1 expresses 
the mean behavior of the low-extinction SNe with the standard deviation of $\sim 0.05$ mag. 
We have used this method to estimate the extinction within the host or the environment around the SN and have tabulated our results in Table 2.

We then compare the derived color excesses to those previously estimated
in the literature. 
We focus on a comparison with the $E(B-V)$ computed by 
Wang et al. (2009b), since these were derived in a 
homogeneous and systematic way (and there is a sufficient 
overlap with our samples), but in Appendix A we also 
compare our values with other estimates of $E(B-V)$ 
found in the literature (reaching the same qualitative conclusion). 
Figure 6 shows a comparison between $E(B-V)$ as derived in this study 
and the values from Wang et al. (2009b), for 16 SNe which are common in 
the two studies. 
Figure 6 shows that the color excess we derived 
tends to be smaller, since a large part of 
the color variations are interpreted to reflect the intrinsic color variations 
rather than the host extinction. 
Indeed, the slope between the two estimates is consistent with $\sim 3 / 1.85$ 
for objects with $E(B-V) \lsim 0.2$, where 1.85 is the value of $R_{V}$ obtained 
by Wang et al. (2009b) for their whole sample. This means that 
even if we assume $R_{V} \sim 3$ to convert our $E (B-V)$ estimate to $A_{\rm V}$, 
the extinctions we derive are mostly consistent with those derived by the other method. 
Exceptions are the heavily reddened SNe, especially SN 2006X. 
This means that at least part of the reason why a small $R_{V}$ is preferred for the 
luminosity standardization might be attributed to an overestimate of the host extinction 
(F10), although at least SN 2006X clearly requires small $R_{V}$. 

\section{Computing Intrinsic Luminosities And Residuals of Type Ia SNe}

\begin{table}
\centering
 \begin{minipage}{140mm}
 \caption{$E(B-V)$}
 \label{tab:sample}
 \begin{tabular}{@{}lcc@{}}
 \hline 
SN & 
$\Delta m_{15} (B)$ Correction\footnote{Corrected for the 
color-$\Delta m_{15} (B)$ relation of F10.} & 
$v_{\rm neb}$ Correction\footnote{Corrected for the 
color-$\Delta m_{15} (B)$ relation {\it and} \\ 
the color-$v_{\rm neb}$ relation (Equation 1).}\\ 
 & (mag) & (mag)\\
\hline
1990N & $0.056$ & $0.030$\\
1994D & $-0.090$ & $-0.019$\\
1997bp & $0.19$ & $0.041$\\
1998aq & $-0.099$ & $-0.079$\\
1998bu & $0.36$ & $0.39$\\
2000cx & $0.14$ & $0.12$\\
2001el & $0.080$ & $0.002$\\
2002bo & $0.45$ & $0.30$\\
2002dj & $0.081$ & $-0.048$\\
2002er & $0.15$ & $0.078$\\
2003du & $-0.063$ & $-0.073$\\
2003hv & $-0.080$ & $0.013$\\
2004dt & $-0.050$ & $-0.010$\\
2004eo & $0.038$ & $0.007$\\
2005cf & $0.003$ & $-0.045$\\
2006X & $1.33$ & $1.23$\\
2006dd & $-0.042$ & $0.000$\\
2007on & $0.082$ & $0.037$\\
2007sr & $0.13$ & $0.020$\\
2009ab & $0.058$ & $0.056$\\
\hline
\end{tabular}
\end{minipage}
\end{table}

To investigate any residuals in the intrinsic luminosities of SNe~Ia, 
it is crucial that the distance to each event is estimated carefully. 
The distance should be provided independently of
the SN properties. 
Most of the SNe Ia in our sample are nearby 
(that is why late-time spectral observations were possible for these objects), 
and therefore the most accurate distance estimates to their host galaxies
come from direct distance measurements such as Cepheid variables or 
the Surface Brightness Fluctuation (SBF) method. 
Although the Cepheid measurements, available for 4 SNe in our 
list, are regarded as the most reliable estimates, there is disagreement 
in the different analyses of the Cepheid data. The Cepheid distance measurements 
have mainly been contributed by the Saha-Tammann-Sandage 
SN Ia Hubble Space Telescope Calibration Program (hereafter STS: 
Saha et al.\ 2006) and 
by the HST key project (hereafter KP: Freedman et al.\ 2001; 
Stetson \& Gibson\ 2001). The KP distances are typically shorter 
by $0.2 - 0.3$ mag than those of STS (see 
Riess et al.\ 2005; Wang et al.\ 2006, for a discussion). 
Unfortunately such a systematic difference is important for our study, 
since it is exactly the order of magnitude of the effect we hope to investigate. 
We have chosen to adopt the KP distances  since they have 
been suggested to give smaller 
dispersion in the residuals (e.g., Wang et al.\ 2006),  and since 
this is the distance scale the SBF distances have been calibrated to (see below). 
In Appendix B, we discuss how our results would be  modified if we  
instead had adopted the  STS distances.

\begin{figure}
   \centering
   \includegraphics[width=0.45\textwidth]{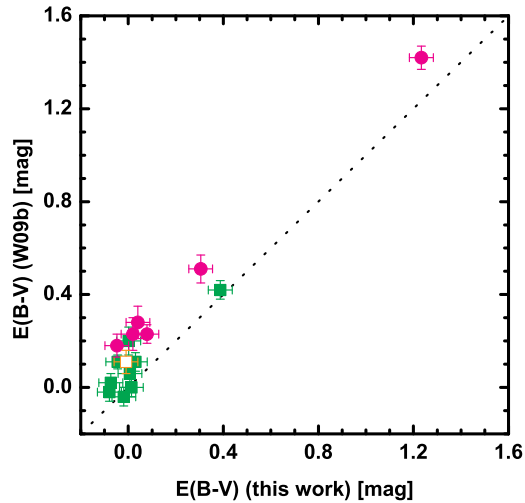}
   \caption{Comparison of the extinction. The horizontal axis denotes the values of 
$E (B-V)$ derived 
in this study corrected for $v_{\rm neb}$ (Table~2), 
while the vertical axis provides values from Wang et al. (2009b). 
The comparison is shown for 16 SNe which are common between the two studies. 
The symbols and color coding indicate 
the velocity gradient near maximum brightness (see Fig. 4 caption).
}
   \label{fig6}
\end{figure}

In the case of  the SBF measurements, we used the values given by Tonry et al. (2001), 
and included the 0.16 mag  correction suggested by 
Jensen et al. (2003) to fit to the KP Cepheid zero-point. 
For the SN host galaxies  not listed in Tonry et al. (2001), we used values given by 
Ajhar et al. (2001) in which the zero-point is calibrated in a similar manner.

There are several objects which require additional considerations about the distance. 
\begin{itemize}
\item For SN 2000cx in NGC 524, 
an SBF distance of $\mu = 31.74 \pm 0.20$ mag is available. 
However, this value is significantly smaller than the Hubble flow distance (see below) 
of $32.63 \pm 0.27$ mag. 
We suspect that the SBF distance to NGC 524 is not correct, 
and have decided to adopt the Hubble flow distance for this SN (see also Appendix B). 
\item Neither a Cepheid nor an SBF distance is available for SN~2001el. We therefore used 
a distance based on the near-IR Tully-Fisher relation (Willick et al.\ 1997). 
\item For SN 2002bo, direct distance measurements are not available. However, 
there are SBF distances measured for two possible members of the 
group (see Krisciunas et al.\ 2004): 
NGC 3193 ($\mu = 32.50 \pm 0.18$ mag) and NGC 3226 
($31.70 \pm 0.24$ mag), which however, do not agree with each other. 
The latter value is consistent with the Hubble flow distance 
($31.44 \pm 0.48$ mag) and we have therefore decided to adopt this latter SBF value. 
\item For SN~2006dd, Feldmeier et al. (2007) reported a distance to NGC~1316 
based on the planetary nebula luminosity function (PNLF) as $\mu = 31.26 \pm 0.10$ mag, which 
Stritzinger et al. (2010) argued is more reliable than the SBF distance ($31.50 \pm 0.17$ mag). 
Therefore, we adopt the PNLF distance in this paper. 
\item In the case of SN~2007sr,  Schweizer et al. (2008) reanalyzed {\em Hubble Space Telescope} 
data used by Saviane et al.\ (2008) and obtained a TRGB 
(tip of the red giant branch) distance of $\mu = 31.51 \pm 0.17$ mag, which is 
in  agreement with the Hubble flow distance. 
We adopt this value as the distance to SN~2007sr. 
\end{itemize}

Finally, there are a number of events for which we had to rely 
on the Hubble flow (HF) distance scale, since estimates based on the aforementioned 
methods are not 
available. Recession velocities were corrected for 
Virgo infall as obtained from the Lyon-Meudon Extragalactic Database 
(Paturel et al.\ 2003), and converted to a distance modulus using  
$H_{0} = 72$ km s$^{-1}$ Mpc$^{-1}$ (Freedman et al.\ 2001). 
The correction with respect to the heliocentric radial velocity 
due to the Virgo infall is typically less than $100$ km s$^{-1}$. 
In these calculations we assumed an uncertainty of 
300 km s$^{-1}$ to account for peculiar motions. 

Reddening due to dust is the other main problem to estimate the SN luminosity. 
Although we suggest that a large part of the color variations is intrinsic 
at least if we omit highly-reddened SNe (\S 4), 
the corrected color excesses are not always negligible (Tab.~2). 
For Galactic extinction we adopt the dust maps of 
Schlegel et al. (1998) with $R_{V} = 3.1$. 
We have estimated the host extinction using the method developed and described in \S 4 
(see Equation 1). 
The associated $E(B-V)$ color excess (Table 2)
was then converted to $A_{V}$ using the relation $A_{V} = R_{V} E(B-V)$. 
The associated error is: 
\begin{equation} 
\sigma_{A_{V}}^2 = R_{V}^2 \sigma_{E(B-V)}^2 + [E(B-V)]^2 \sigma_{R_{V}}^2 \ .
\end{equation}
Estimating the uncertainty in $R_{V}$ is not trivial. We have assumed the typical 
value derived by F10, i.e. $\sigma_{R_{V}} = 0.3$. 
Following the procedure of the previous section, we adopted an error of $0.05$ mag 
for $E(B-V)$. 
We assumed $R_{V} = 1.72$ since to derive the residuals we used the 
light curve correction from F10, 
who obtained this value as the best fitting parameter. 
In Appendix B, the effect of 
the uncertainty in $R_{V}$ upon our results is further investigated. 

We note that there are two approaches in treating the color and extinction for SNe Ia 
(see, e.g., F10 for further discussion). 
In one approach, the color and the light curve shape are simultaneously varied to 
obtain the standardized luminosity (e.g., Guy et al. 2007). 
In the other approach, 
first the intrinsic color is associated with other observed parameter(s) (e.g., $\Delta m_{15} (B)$) 
and any excess 
beyond this color is assumed to be caused by the host extinction (e.g., Phillips et al. 1999). 
We follow the latter approach in this paper. We however emphasize that our treatment takes into 
account the intrinsic color variations independent from $\Delta m_{15} (B)$, as is done 
in the former, simultaneous-fit approach. 

\begin{table}
\centering
 \begin{minipage}{140mm}
 \caption{Peak Luminosity Calibration}
 \label{tab:luminosity}
 \scriptsize
 \begin{tabular}{@{}lccccc@{}}
 \hline 
SN & 
$\bar M_{V} (\Delta m_{15} (B))$ & 
$M_{V}$ & $dm$ & 
$\sigma_{A_{V}}$\footnote{Error in the extinction estimate.} & 
$\sigma_{M_{V}}$ \footnote{Total error including the extinction and distance 
uncertainties.} \\ 
& (mag) & (mag) & (mag) & (mag) & (mag)\\
\hline
1990N & $-19.15$ & $-19.14$ & $0.01$ & $0.087$ & $0.17$ \\
1994D & $-18.91$ & $-19.12$ & $-0.21$ & $0.086$ & $0.22$ \\
1997bp & $-19.24$ & $-18.97$ & $0.27$ & $0.087$ & $0.28$ \\
1998aq & $-19.10$ & $-19.01$ & $0.09$ & $0.089$ & $0.12$ \\
1998bu & $-19.16$ & $-18.97$ & $0.19$ & $0.14$ & $0.25$ \\
2000cx & $-19.28$ & $-19.85$ & $-0.57$ & $0.093$ & $0.29$ \\
2001el & $-19.09$ & $-18.55$ & $0.54$ & $0.086$ & $0.46$ \\
2002bo & $-19.06$ & $-18.72$ & $0.34$ & $0.13$ & $0.27$ \\
2002dj & $-19.14$ & $-18.89$ & $0.25$ & $0.087$ & $0.26$ \\
2002er & $-18.91$ & $-18.91$ & $0.00$ & $0.089$ & $0.27$ \\
2003du & $-19.20$ & $-18.76$ & $0.44$ & $0.089$ & $0.31$ \\
2003hv & $-18.64$ & $-18.90$ & $-0.26$ & $0.086$ & $0.31$ \\
2004dt & $-19.02$ & $-19.28$ & $-0.26$ & $0.086$ & $0.15$ \\
2004eo & $-18.79$ & $-19.11$ & $-0.32$ & $0.086$ & $0.16$ \\
2005cf & $-19.10$ & $-18.84$ & $0.26$ & $0.087$ & $0.34$ \\
2006X & $-18.92$ & $-19.07$ & $-0.15$ & $0.38$ & $0.43$ \\
2006dd & $-19.14$ & $-18.94$ & $0.20$ & $0.086$ & $0.13$\\
2007on & $-18.69$ & $-18.54$ & $0.15$ & $0.087$ & $0.21$ \\
2007sr & $-19.12$ & $-19.01$ & $0.11$ & $0.086$ & $0.19$\\
2009ab & $-18.98$ & $-18.79$ & $0.19$ & $0.088$ & $0.22$\\
\hline
\end{tabular}
\end{minipage}
\end{table}

The observed $V$-band peak magnitude was converted to absolute peak 
magnitude ($M_{V}$) 
using the distance modulus and the extinction as described above. 
We have not included K-corrections, 
since the correction is at most $\sim 0.01$ mag at maximum 
brightness for SNe at the low redshifts considered here 
(Hamuy et al.\ 1993; Nugent et al.\ 2002). 
For the residual, the difference between the absolute peak $V$ magnitude 
($M_{V}$) and the standardized magnitude 
predicted by the Phillips relation [$\bar M_V (\Delta m_{15} (B))$] 
was computed. 
In obtaining the ``standardized'' peak magnitude, 
we applied an updated relation presented by F10 (see their Table~9.). 
Hereafter, we denote the residual by $dm \equiv 
M_{V} - \bar M_{V} (\Delta m_{15} (B))$. 
Table 3 summarizes the $M_{V}$ values that we computed, 
along with the standardized peak magnitudes. 
Also shown are the estimated errors in the calibration. 

In the resulting distribution of $M_{V}$, 
the dispersion is at the level of $\sim 0.28$ mag. 
This value is larger than that typically derived from samples of SNe~Ia 
($\sim 0.1 - 0.2$ mag; e.g., F10). 
We note, however, that such studies make use of uniform data sets and 
minimize the dispersion by parameter fitting the entire data set, 
including simultaneous fit to 
light-curve shape (e.g., $\Delta m_{15} (B)$) and color (i.e., $R_{V}$ and intrinsic SN color). 
Given that we have compiled data from various sources and adopted an 
external relation (F10) derived from an independent data set, 
a larger dispersion is expected. 
More importantly, 
most of the SNe~Ia in our analysis are nearby and are not in the smooth Hubble flow, 
so that the uncertainty in the distance estimates is larger (due to 
peculiar velocities; see for example, fig.~19 of F10).

\section{Relation Between The Late-Time Emission Line Shift And The Luminosity Residuals}

\begin{figure}
   \centering
   \includegraphics[width=0.45\textwidth]{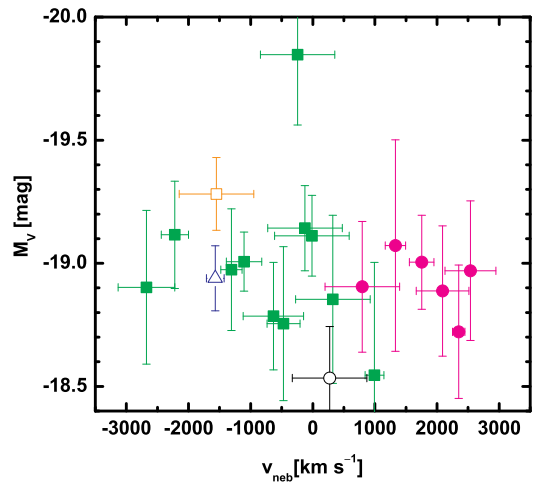}
   \caption{$M_{V}$ as a function of $v_{\rm neb}$. 
The symbols and color coding indicate 
the velocity gradient near maximum brightness (see 
Fig. 4 caption). }
   \label{fig7}
\end{figure}

In this section we explore whether there are any relations 
between $v_{\rm neb}$ and the luminosity of SNe~Ia
at maximum brightness. 
We have first investigated whether $v_{\rm neb}$ correlates with the absolute magnitude 
$M_{V}$. 
In Fig.~7 it is shown that there is no apparent
correlation between these two parameters. 
The observed distribution of $M_{V}$ as a function of $v_{\rm neb}$ 
can emerge from a non-correlation 
at the 14\% level ($P = 0.14$), 
as derived by our Monte-Carlo simulations (see Appendix C). 
There is a clear outlier in the plot, that is SN 2000cx 
(see also Li et al. 2001). 
Even if we omit this single SN from the fit, the fitting result 
is not improved ($P = 0.19$).

\begin{figure}
   \centering
   \includegraphics[width=0.45\textwidth]{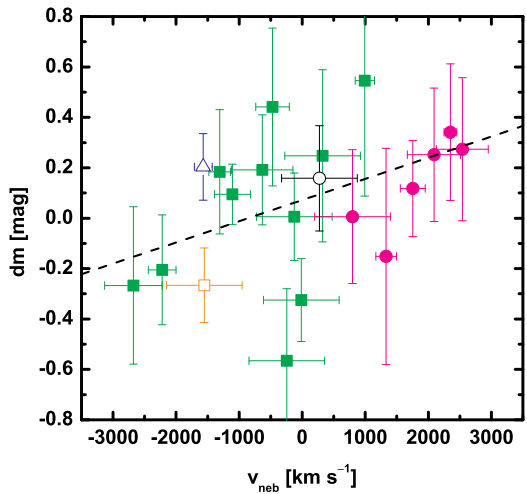}
   \caption{
The residuals $dm$ as a function of $v_{\rm neb}$ (see Figure 4 caption). 
The black dashed line shows the best linear fit to the entire data set (Tab.~5). 
}
   \label{fig8}
\end{figure}

\begin{table*}
\centering
 \begin{minipage}{140mm}
 \caption{Relations between $v_{\rm neb}$ and $dm$}
 \label{tab:mvvneb}
 \begin{tabular}{@{}ccrrcc@{}}
 \hline 
sample & N\footnote{Number of SNe.} & 
$\alpha$\footnote{The best fit using the relation 
$dm = \alpha (v_{\rm neb}/1000 {\rm \ km \ s}^{-1}) + \beta$. 
The errors for $\alpha$ and $\beta$ are $1\sigma$.} & 
$\beta$ & 
$P$\footnote{Probability that the distribution arises from a non-correlation.} & 
significance\\
\hline
 All & 20 & $0.084 \pm 0.051$ & $0.072 \pm 0.077$ & $0.053$ & $1.6\sigma$\\
  `Bright'\footnote{SNe with $\Delta m_{15} (B) \le 1.12$.} & 
10 & $0.028 \pm 0.071$ & $0.12 \pm 0.10$ & $0.35$ & $0.4\sigma$\\
  `Faint'\footnote{SNe with $\Delta m_{15} (B) \ge 1.13$.} & 
10 & $0.13 \pm 0.079$ & $0.056 \pm 0.13$ & $0.054$ & $1.6\sigma$\\\hline
\end{tabular}
\end{minipage}
\end{table*}

In the next step, we investigated 
if $v_{\rm neb}$ correlates with the luminosity residual $dm$ (Fig.~8). 
According to the Monte-Carlo simulations (see Appendix C), 
these two quantities are correlated at $1.6\sigma$ level. 
The best linear fit to the 20 data points, expressed by 
$dm = \alpha \ (v_{\rm neb}/1000~{\rm km~s}^{-1}) + \beta$, is given in Table 4. 
We note that this investigation is 
limited by the relatively large errors in $dm$ associated with the distance uncertainties 
to the SNe (\S 5 and Appendix B).
If we arbitrarily choose to ignore these errors (that 
were computed with a range of different methods) and consider only the errors
associated with the extinction (that were derived in a systematic way), we observe that the significance of the proposed correlation increases ($2.4\sigma; P = 0.0092$). 
Although the significance for the observed correlation is low, 
it deserves further study 
when a larger sample is available. 

There may be a concern that the apparent relation in the 
$v_{\rm neb} - dm$ could be dominated by single points, rather than expressing a general trend. 
To examine this, we artificially removed single points from the data set, 
and determined how much the quality of the fit was affected. We thereby obtained 
$P = [0.038 - 0.086]$ 
($1.4\sigma - 1.8\sigma$), depending on which of the 20 points was removed. 
Also, the resulting fits were all consistent within the $1\sigma$ error. 

We also split the SN sample into halves according to the value of $\Delta m_{15} (B)$, and 
repeated the fitting for both subsamples (Table 4). 
We find that the `faint' group (with $\Delta m_{15} (B) \ge 1.13$ mag\footnote[7]{Here the dividing 
value, $\Delta m_{15} (B) = 1.13$ is set merely by the requirement to split the sample into  halves.}) 
shows a correlation with a significance similar to that using the entire sample. 
On the other hand, 
the correlation is not evident for `bright' SNe with $\Delta m_{15} (B) \le 1.12$. 
We note that the best fit slope for the faint sample is steeper than the one for the entire sample, as expected from the behavior of the bright sample.
A possible interpretation is presented in \S 7.

\begin{figure}
   \centering   
  \includegraphics[width=0.45\textwidth]{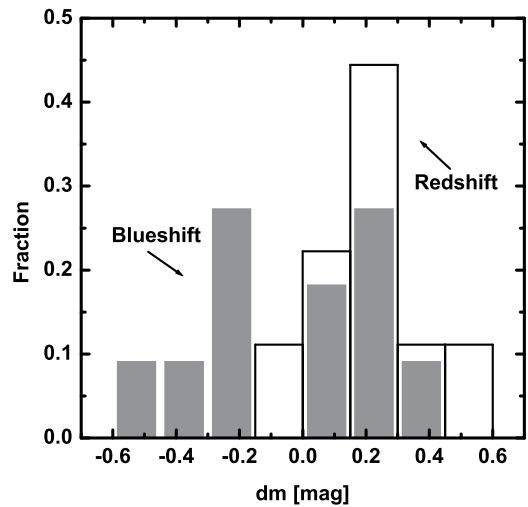}
   \caption{Fractional number distribution of the 20 SNe Ia 
as a function of $dm$.
SNe~Ia are divided into two categories: 
those showing a blueshift in the late-time emission lines (gray) 
and those showing a redshift (white).}
   \label{fig9}
\end{figure}

We have also investigated a tendency by dividing the sample depending on 
$v_{\rm neb}$. Figure 9 shows the number fraction of SNe Ia 
as a function of $dm$. In this figure, the sub-sample with $v_{\rm neb} < 0$ 
(blueshift) and that with $v_{\rm neb} > 0$ (redshift) are marked differently. 
Fig.~9 shows that there is a 
tendency that SNe Ia having 
blueshifts/redshifts in the late-time emission lines are brighter/fainter 
than expected from the Phillips relation. 

The difference in the average values for $dm$ in the `blueshift' group 
($N = 11$) and `redshift' group ($N = 9$) is $\sim 0.25$ mag.  
To examine the statistical significance, we performed Monte-Carlo simulations 
as follows. First, we randomly selected 11 out of the 20 SNe (``A group"), 
and then the remaining 9 SNe were labeled as ``B group".  We then calculated 
the difference in the average values of $dm$ in the two groups. 
With $10^{5}$ Monte-Carlo realizations, 
we find that the probability, $P$, that $dm$ (A) $-$ $dm$ (B) $\lsim -0.25$ mag, 
is $P = 0.023$ ($2.0\sigma$). 

\section{A Toy Model for The Viewing Angle Effects}

In the off-centre asymmetric SN Ia explosion scenario (\S 2), 
a correlation between $v_{\rm neb}$ and $dm$ is expected. 
The viewing angle effect on the peak brightness has been investigated by several authors. 
Sim et al. (2007) considered kinematic models for
an off-centre distribution of $^{56}$Ni, computed the radiation transfer in 2D, 
and showed that the luminosity can be different by 
$\sim 0.5 - 1.5$ mag for an observer at the offset direction as compared to 
the opposite direction. More recently, Kasen et al. (2009) investigated  
a series of off-centre hydrodynamic models in 2D, in which the deflagration was 
ignited in an off-centre manner (see also M10c), 
obtaining peak magnitude differences at the level of 
$\sim 0.5$ mag for different viewing angles. 
There is also an expectation that the color is affected by the viewing angles, as 
investigated by Kasen et al. (2009) and Foley \& Kasen (2010). 
However, there was no strong observational 
hint on the geometry of the inner ejecta 
at the time of their publications (but see Foley \& Kasen 2010). 
In addition, they did not discuss the outcome of the off-centre ignition 
at the late phases as was investigated by M10ab. 
In the following, we investigate the effect of the viewing angle 
on the peak luminosity and color, adopting the simple explosion geometry proposed by 
M10ab. 

\begin{figure}
   \centering
   \includegraphics[width=0.45\textwidth]{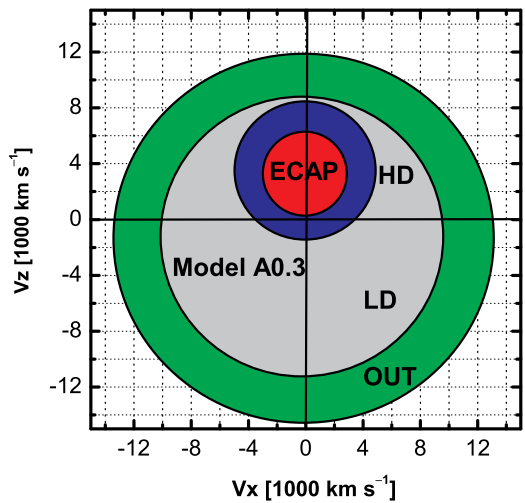}
   \caption{The toy model used for the light curve calculations. The structure of 
the ECAP and HD zones is directly taken from the model of SN 2003hv (M10a). 
}
   \label{fig10}
\end{figure}

In the toy model shown in Fig.~10, 
the innermost regions are divided into two characteristic zones: 
a zone dominated by neutron-rich Fe-peak elements produced by electron captures 
(ECAP; Electron-CAPture), the relatively high density 
$^{56}$Ni-rich region surrounding the 
ECAP region (HD; High-Density), and then the outer, relatively low-density, 
$^{56}$Ni-rich region on top of the HD region (LD; Low-Density). 
The ECAP and HD regions are suggested to be responsible for 
the formation of [Ni II]~$\lambda$7378 and [Fe II]~$\lambda$7155, respectively, 
from which we derived $v_{\rm neb}$. 
Following M10ab, 
the ECAP and HD regions have offsets from the centre of $3,500$ km s$^{-1}$, 
whereas the LD region has only a slight offset in the 
opposite direction. 
The ECAP and HD zones 
are interpreted as products of the initial deflagration flame propagation, 
while the LD zone is a product of the subsequent detonation wave. 
There is possibly a variation of the offset velocity for different SNe Ia, 
although in the present analysis we use one specific value ($3,500$ km s$^{-1}$) 
as our fiducial model.\footnote[8]{The presently available data analyzed by M10ab 
are consistent with no variation. To tackle the possible variation in the 
kinematics and its relation to other explosion parameters would require 
a larger observational sample. } 
Note that this distribution was derived by modeling late-time spectra, 
but it is also qualitatively consistent with the outcome of hydrodynamic models if 
the deflagration sparks are ignited at an offset (M10c). 
The hydrodynamic simulations have not yet explored all parameter space,
nor been directly compared to observations. We therefore take 
this `observationally constrained' toy model as our reference model. 

The masses and compositions of the ECAP and HD zones are taken from 
M10a; the mass of $^{56}$Ni is $0.01~M_{\odot}$ in the ECAP 
zone and $0.1~M_{\odot}$ in the HD zone. 
The mass of $^{56}$Ni in the LD zone is a parameter, corresponding to the diversity 
of $M$($^{56}$Ni) and $\Delta m_{15} (B)$ for various SNe Ia. 
The radial extent of the LD zone is also taken from M10a. 
We calculate light curves for models with total $M$($^{56}$Ni) 
from $0.21$ to $0.8~M_{\odot}$, with 10\% increases in 
$M$($^{56}$Ni) between models. In the present analysis, we focus on the models with 
$M$ ($^{56}$Ni) $= 0.3~M_{\odot}$ 
(roughly consistent with the $0.3 - 0.4~M_{\odot}$ derived for  
SN 2003hv; Leloudas et al. 2009; M10a) 
and $0.6~M_{\odot}$ (for SNe~Ia with a typical luminosity; see Stritzinger et al.\ 2006). 
We hereafter refer to the models as 
A0.3, A0.6, where the number denotes $M$($^{56}$Ni). 
For all the models, the structure in the ECAP and HD zones (i.e., the 
masses of these zones and the offsets) was assumed to be exactly the same as the one derived for SN 2003hv. 
We fixed the masses in the ECAP and HD zones since it has been suggested  through late-time spectral modeling (Mazzali et al. 2007) that 
the masses of the innermost region are 
consistent with no variation for different SNe. 

The distribution of elements other than Fe-peak elements was not discussed 
in M10a, because a large fraction of them are located in 
the region above the LD-zone which is not accessible by late-time 
observations. 
These are assumed to be intermediate mass elements (IMEs) 
and/or a carbon-oxygen (CO) mixture. 
We assume an explosion energy of $E_{\rm K} = 1.4 \times 10^{51}$ erg
and ejected mass of $M_{\rm ej} = 1.38~M_{\odot}$.\footnote[9]{Strictly speaking, 
$E_{\rm K}$ is dependent on the final composition. If $M$($^{56}$Ni) $= 0.6~M_{\odot}$ and 
the remaining part is burned into intermediate mass elements, then $E_{\rm K} \sim 1.4 
\times 10^{51}$ erg. By varying $M$($^{56}$Ni) between $0.2$ and $0.8~M_{\odot}$ and 
assuming that the remaining fraction is burned into IMEs, 
the resulting $E_{\rm K}$ is in the range $(1.3 - 1.5) \times 10^{51}$ erg. Thus, the variation 
in $E_{\rm K}$ is small, and we take $E_{\rm K} = 1.4 \times 10^{51}$ erg 
as our fiducial value.} These values determine the outermost velocity, 
as well as the density there. IMEs and CO are assumed to be distributed uniformly in the 
LD and the outer zones. Accordingly, the mass in the LD zone is 
the sum of those of $^{56}$Ni and IMEs/CO, whereas the outermost zone is 
purely composed of IMEs or CO. 

A question is whether or not the outermost region also shows an offset. 
To address this, we employ a simple model.  
It is assumed that the centre of the outer envelope is 
the same as for the LD-zone. This is motivated by the expectation that 
the outer envelope expansion is caused by the detonation wave. 
This configuration is roughly consistent with the finding by M10b that 
SNe viewed from the offset direction (direction of the ECAP zone) are observed 
as LVG SNe while those from the opposite direction are HVG SNe because 
of the existence of a more extended envelope in the direction opposite to the offset.  
This characteristic is also seen in a hydrodynamic explosion model (see M10b 
for a detailed discussion). 

\begin{figure} 
   \centering 
   \includegraphics[width=0.45\textwidth]{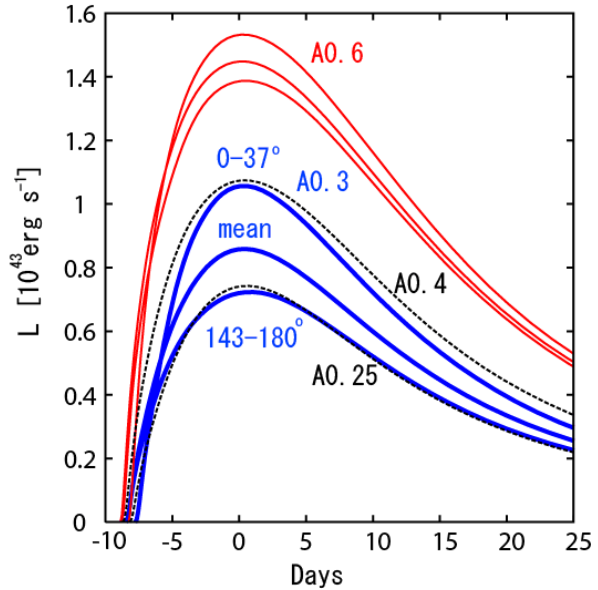} 
   \caption{The synthetic bolometric light curves for models A0.3 (blue-thick-solid lines) 
and A0.6 (red-thin-solid lines). For each model, three curves are shown: 
Two angle-dependent light curves, 
with the angle measured from the 
direction of the offset of the ECAP/HD regions, and the mean angle-averaged 
light curve. 
For comparison, the mean light curves for models A0.25 and A0.4 
are shown (black-dashed lines). 
Time is defined as days past bolometric maximum.
}
   \label{fig11}
\end{figure}

For different masses of $^{56}$Ni, we simulated the bolometric light curve 
as a function of the viewing angle. 
The calculations were performed 
using the Monte-Carlo radiation transfer {\it SAMURAI} code. 
{\it SAMURAI} is a compilation of 3D codes that adopt Monte-Carlo methods 
to compute high-energy light curves and spectra (Maeda\ 2006a), optical 
bolometric light curves (Maeda et al.\ 2006b), and optical spectra from early 
(Tanaka et al.\ 2006, 2007) to late phases (Maeda et al.\ 2006c). 
For the light curve, the present version allows for a frequency-averaged 
gray opacity only. The optical transfer scheme 
follows prescriptions given by Lucy (2005) (see also Cappellaro et al. 1997). 
For the opacity to optical photons, we adopted a phenomenological prescription of 
Mazzali et al. (2001)\footnote[10]{We note that the rise time in the models is typically
shorter than that estimated by observations. 
A part of the reason is likely the simplified opacity used in our calculations (see also 
Sim et al.\ 2007). As we are only interested in the 
`relative' behavior, e.g., differences caused by different 
viewing angles and by different $M$($^{56}$Ni), we did not try to obtain better agreement 
in the rise time.}. Because of the simplified kinematic model and the limitations in the 
light curve calculations, the following results should be regarded as indicative. 
A more detailed and quantitative comparison between model and observations is 
beyond the scope of this paper.

Figure 11 shows the bolometric light curves of models A0.3 and A0.6, for different viewing angles. 
Model A0.3 shows a difference of $\sim 30$\% in the peak luminosity depending on the 
viewing angle. It is brightest if viewed from the offset direction, and faintest 
from the opposite direction. The reason for this is that 
photons from the HD zone have lower optical depth in the offset direction 
and preferentially escape into this direction. 
On the other hand, the contribution from the nearly spherical LD component is not very 
sensitive to the viewing angle. In model A0.3, a large fraction of $^{56}$Ni ($\sim 30$\%) is 
contained in the offset component. The brightness thus depends strongly on 
the viewing angle. On the other hand, in model A0.6, 
the contribution of the offset component to $M$($^{56}$Ni) 
is small ($\sim 15$\%). Accordingly, the angle-dependence is small, at the $\sim 10$\% level. 

Another important effect is that the light-curve shape can also be 
different for different viewing angles. Model A0.3 shows that the light curve is narrower and 
the rise time is shorter if viewed from the direction of the offset, while it 
is broader and the rise time is longer if viewed from the opposite direction (see also 
Sim et al.\ 2007). 
This stems from the shorter diffusion time scale in the direction of the offset. 
Indeed, this effect is part of the reason for the enhancement of the brightness in 
the offset direction, since faster rise to the peak results in a larger amount of 
radioactive $\gamma$-rays available to power the peak brightness. 
The presently available data do not reveal this theoretically expected 
correlation between $\Delta m_{15} (B)$ and $v_{\rm neb}$ (M10b), 
although the situation is still consistent 
with the idea presented here 
since a large variation in $\Delta m_{15} (B)$ is introduced by different amounts of 
$^{56}$Ni. With an increasing number of nebular spectra, this test 
will become possible. 

\begin{figure}
   \centering
   \includegraphics[width=0.45\textwidth]{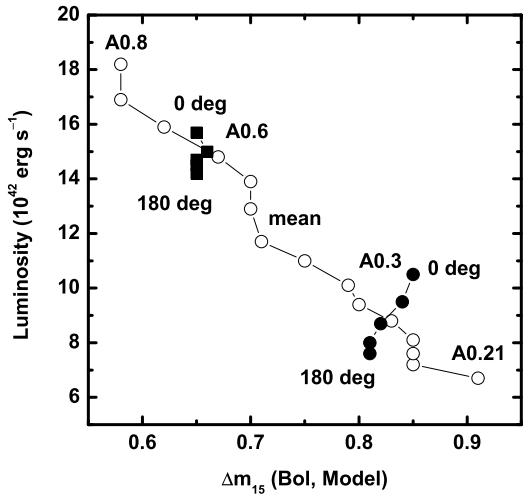}
   \caption{The behavior of the light curve shape and the peak luminosity of our models. 
The horizontal axis denotes $\Delta m_{15}$ in the synthetic bolometric 
light curves, and the vertical axis the synthetic bolometric luminosity. 
Open circles denote angle averaged, mean light curves of different models 
(from A0.21 to A0.8). The angle-dependent behavior is shown for models A0.3 and 
A0.6 by black filled symbols, where each point stands for the light curve from a 
certain direction. }
   \label{fig12}
\end{figure}

The trend in the peak brightness -- light-curve width relation arising from different viewing angles 
works in a way opposite to the Phillips relation (see also Sim et al. 2007). 
A light curve that is brighter due to asymmetry is also narrower. 
For example, Fig.~11 shows that the peak brightness of the 
model A0.3 viewed from the offset direction is comparable to the mean (angle-averaged) 
brightness of model A0.4, while the former results in a narrower light curve. 
Again, this effect is not as important for model A0.6, 
for which the light curve shape 
does not strongly depend on the viewing direction. 
Figure 12 shows the peak luminosity in terms of $\Delta m_{15}$ as derived from 
the synthetic bolometric 
light curves.\footnote[11]{Note again that our simplified treatment in the opacity and in the ejecta geometry allows us to investigate only the relative behavior of different 
models (and viewing directions), but not to directly compare the models and the 
observational data. }

Figure 13 shows the synthetic bolometric magnitude `difference'  ($\sim dm$)
as a function of $v_{\rm neb}$ (i.e., the viewing angle). 
It is compared to the $V$-band 
residuals as derived in the previous section. Given the uncertainty 
involved in the comparison (e.g., the simplified ejecta structure), 
it should be regarded as qualitative.

\begin{figure}
   \centering
   \includegraphics[width=0.45\textwidth]{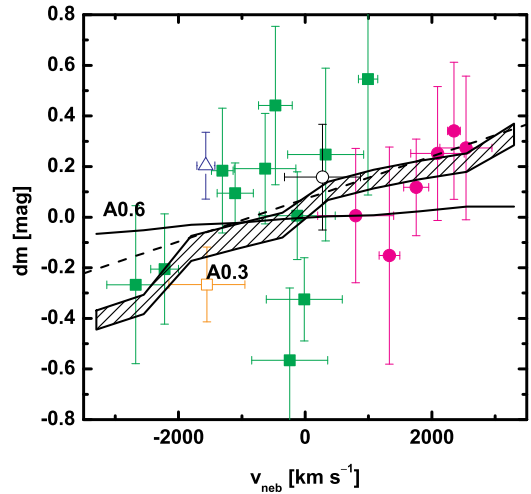}
   \caption{Model results compared to the observed sample in the 
$v_{\rm neb}$ vs. $dm$  diagram.  
Model A0.3 is shown by the shaded area, 
model A0.6 by a solid line (see footnote 12). 
Points are for our sample of SNe, 
and a dashed line is the best linear fit to them (Fig.~8). 
}
   \label{fig13}
\end{figure}

The model behavior in Figure 13 was derived through the following procedures. 
First, we defined a bolometric Phillips relation for the models, 
based on the angle-averaged, mean light curves in Figure 12. 
Then, for a light curve of model A0.3 and {\em a given viewing direction}, 
we searched for a `mean' light curve (spanning from model A0.21 to model A0.8) 
which shows a similar light curve shape. 
In general, the considered light curve lies between two 
mean curves of our model grid.\footnote[12]{For model 
A0.6, the light curve shape is not strongly dependent on 
the viewing direction (Fig. 11). Thus, for model A0.6, the `model residual' is computed 
as the magnitude difference between the angle-dependent and mean light curves of the 
same model A0.6. This is the reason why in Figure 13 
the model residual is expressed by a single line 
for model A0.6, unlike model A0.3.}. 
Next, we define the model `residual' as the magnitude difference between the 
{\em angle-dependent} luminosity under consideration and the luminosity 
of the corresponding mean light curve(s). 
This procedure is equivalent to 
deriving the observed residual ($dm$), but with the observed reference light curves 
replaced by the synthetic `mean' (angle-averaged) light curves. 

To compare the model residual (in bolometric magnitude) and observed residual $dm$ (in 
the $V$ band), 
we have to make one assumption: The $V$-band light curve is (at least qualitatively) 
traced by the bolometric light curve. 
Using a sample of SNe with the bolometric luminosity available (Contardo et al. 2000; 
Stritzinger \& Leibundgut 2005), 
we confirmed that $V_{\rm max}$ is strongly correlated with the bolometric luminosity, 
and that the bolometric correction is not significantly different for different SNe. 
This justifies our assumption. 

In this comparison, not only can the dependence of the peak brightness on the 
viewing angle be important but also that of the light-curve shape. 
For instance, the residual for model A0.3 viewed from the offset direction 
should be constructed using the mean (angle-averaged) peak brightness of A0.23 or A0.25, 
which have similar light curve shapes, but smaller luminosity than, model A0.3. 
A similar argument applies to an observer 
in the direction opposite to the offset direction. 
Thus, the resulting model $dm$ is generally larger 
than that estimated by simply taking the difference between the angle-dependent brightness 
and the mean brightness for the {\it same model}. 
For example, the residual for model A0.3 is $\sim -0.4$ mag if viewed from the offset direction, 
$\sim +0.3$ mag if viewed from the opposite direction. 
Thus the residual varies by up to $\sim 90$ \% ($\sim 0.7$ mag) 
depending on the viewing angle. 
This is much larger than simply taking the difference between the angle-dependent brightness 
and the mean brightness for the {\it same model}, which is $\sim 30$\% (Fig.~11). 

There is a qualitative agreement between the model and the data (Fig. 13). 
In particular, models with $M$($^{56}$Ni) $\sim 0.3~M_{\odot}$ (model A0.3) 
agree with the tentative correlation seen in the observed data. 
Also, if we calculate the expected difference in $dm$ for SNe with blueshift (observed at 
$< 90^{\circ}$) and redshift ($> 90^{\circ}$), 
taking into account the solid angle, we obtain $\sim 0.4$ mag for this model. 
This is a bit larger than the observed difference in the two groups 
(Fig.~9 and related discussion), but it explains the observed trend 
at least qualitatively. 
Note that the offset velocity has been directly constrained for model A0.3 using SN 2003hv (M10a). 

For brighter SNe Ia, model A0.6 predicts a peak brightness almost independent 
of the viewing angle. We also find that the viewing angle effect on the light curve 
shape is not as important for this model. 
The reason for this is that most of the $^{56}$Ni is in the 
`spherical' LD zone in these models, since the amount of $^{56}$Ni 
in the ECAP and HD zones (the ``offset'' region) is fixed. 
Our analysis of the observational data suggests that 
the correlation is indeed weaker,
as quantified by the correlation slope, 
for a subset of (brighter) 
SNe with smaller $\Delta m_{15} (B)$ than 
for those with larger $\Delta m_{15} (B)$ (see Table 4 and related discussion). 
This is consistent with the model prediction, and may indicate that the contribution 
from the LD zone is more important and the viewing angle effect less important 
for brighter SNe. 
Although we fixed the mass of the offset component in our toy models, 
there is a theoretical expectation that 
the mass of the ``offset'' region is {\it smaller} for brighter SNe, 
since a larger asymmetry results in a weaker deflagration which is then followed by 
a stronger detonation in the off-centre deflagration-to-detonation transition 
scenario (Kasen et al. 2009; M10c). This will result in 
even smaller angle dependence for typical SNe Ia than model A0.6 predicts. 
A larger sample will be required to 
clarify the relative contributions from the different zones, which in turn 
will provide hints on the geometry of SNe as a function of $\Delta m_{15} (B)$. 

\begin{figure}
   \centering
   \includegraphics[width=0.45\textwidth]{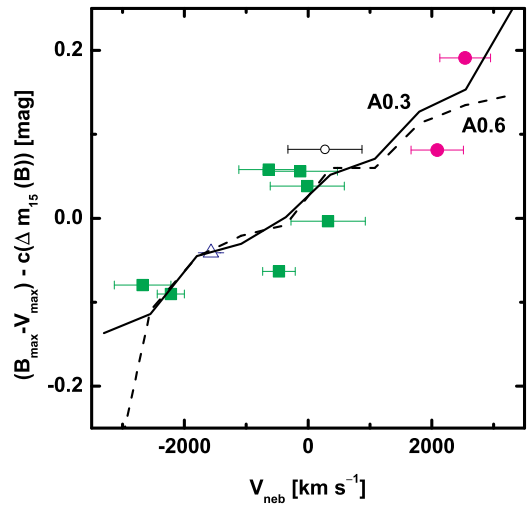}
   \caption{Model results compared to the observed sample in the 
$v_{\rm neb}$ vs. $B_{\rm max}-V_{\rm max}$ diagram (see the main text).  
Models A0.3 and A0.6 are shown by solid and dashed lines, 
respectively. 
Points are for our low-extinction sample of SNe (Fig.~4a). 
}
   \label{fig14}
\end{figure}

Having computed angle-dependent luminosities, it is now possible to 
make a crude estimate on the relation between the color and 
the viewing direction for our models. The angle dependent effective temperature 
is approximately extracted from the model, by combining the luminosity and 
photospheric velocity obtained by integrating the optical depth back inward 
along each radial direction (i.e., $T_{\rm eff} \propto v_{\rm ph}^{-1/2} L^{1/4}$ 
where $v_{\rm ph}$ and the luminosity $L$ are angle dependent). 
Then we convert the effective temperature to the maximum color $B-V$ 
assuming a relation between these two obtained for synthetic spectra 
by Nugent et al. (1995) (Their figure 3).  Since the observational data (Fig. 4) 
are already corrected for $\Delta m_{15} (B)$, we add a constant to the model 
$B-V$ color (i.e., shifted vertically) so that it 
passes $B-V \sim 0$ at $v_{\rm neb} \sim 0$ (i.e., 
we focus on relative color change depending on the viewing direction). 
The result is shown in Figure 14. 

Since the model prediction will be sensitive to the assumed outer ejecta structure, 
this comparison should be regarded to be qualitative. However, this does 
demonstrate that the basic features in the ejecta structure constrained by the 
observations (the inner structure from the late-time spectra and the outer structure 
from the relation between the early-time and late-time spectra; M10ab) 
predict the $v_{\rm neb}$ - color relation found in this paper as well, although 
the models have not been tuned at all to reproduce this relation. 
Indeed, Foley \& Kasen (2010) have recently reported the same trend 
in the color and the viewing direction in their simulation based on 
a qualitatively similar ejecta structure, independently from this 
study. The qualitative behavior, i.e., bluer color for the direction closer to the offset, 
stems from the lower expansion velocity, thus higher temperature, in that direction. 

Unlike the bolometric luminosity, the relation is insensitive to the mass of $^{56}$Ni, 
since the temperature is only weakly dependent on the luminosity. Under the 
assumptions within our toy model, the two models - A0.3 and A0.6 - do 
predict almost the same relation. This may indeed support why we see 
in the observations a strong relation in the color but not in the luminosity residual. 

\section{Discussion and Conclusions}

In this paper, we have investigated how 
the explosion geometry and viewing angle can influence 
the color and peak brightness of SNe Ia, and thereby lead to the residuals
that remain  in the peak magnitudes after application of the light-curve correction. 
We have used the wavelength shift of late-time emission lines 
to derive the viewing angle, and then compared this quantity with the 
color and the luminosity residual for a sample of 20 SNe Ia. 

\subsection{Intrinsic Color Variation}
  
We have found a correlation between the color at maximum and 
the wavelength shift seen in late-time emission lines. 
Using a sub-sample of 11 SNe which likely suffer from insignificant host reddening, 
selected based on the morphological type of the host galaxy and the position of the SN within the host, 
we have found that SNe which show  a blueshift/redshift in nebular emission lines 
(i.e., viewed from the offset/anti-offset direction in the off-centre ignition scenario) 
are bluer/redder than expected from the color-$\Delta m_{15} (B)$ relation. 
This indicates that the previously suggested color difference 
between the `high-velocity-gradient' (HVG) SNe and `low-velocity-gradient' (LVG) SNe 
is connected to the viewing direction to the observer. 

In a next step, we expanded our sample, including objects with potentially larger host reddening. 
Except for 3 very red SNe [($B_{\rm max} - V_{\rm max}) > 0.2$ mag: SNe 1998bu, 2002bo and 2006X], 
all other objects agreed well with the color vs. $v_{\rm neb}$ relation derived from the 
``low-extinction'' sample. 
This indicates that a significant part of the color excess previously attributed 
to host extinction is actually due to intrinsic color variations. 

This raises the question whether the often preferred value of $R_{V} \lsim 2$ 
(much smaller than the typical Galactic value of $\sim 3.1$) really reflects 
different properties of interstellar/circumstellar dust around SNe Ia. 
F10 argued that the optical color minus NIR color relation 
can be reproduced even with $R_{V} \sim 3.2$ once heavily reddened SNe are omitted. 
On the other hand, they required $R_{V} \sim 1 - 2$ 
to minimize the dispersion in the standardized luminosity calibration. 
They pointed out that an intrinsic color variation which does not correlate with the decline 
rate, but does correlate with the luminosity, 
may solve this apparent discrepancy. Indeed, the intrinsic variation related to the viewing angle 
that we have found in this paper does not correlate with the decline rate. 
If we exclude SNe which are clearly heavily reddened, we see that $R_{V}$ as large as 
the Galactic value could be acceptable. 
This issue requires more careful and systematic analysis 
based on a larger sample. Also, we note that even if this works for the majority of SNe Ia, 
some heavily reddened SNe appear to require $R_V \sim 1 - 2$, 
as highlighted by SN 2006X (F10). 

A toy model constructed with the constraints so that it explains the late-time spectral 
variation (M10a) and its relation to the early-phase spectral diversity (M10b) predicts the bluer/redder 
color for smaller/larger (blueshift/redshift) $v_{\rm neb}$, as is consistent with the observational data. 
For our model sequence, the predicted trend is not sensitive to $M$($^{56}$Ni) (or $\Delta m_{15} (B)$). 

\subsection{Second Parameter in the Luminosity Calibration?} 

We have also investigated a relation 
between the SN Ia luminosity residual after calibration with the Phillips relation, 
and the wavelength shift seen in the late-time nebular emission lines. 
For our small sample, the correlation is not strong, and we regard 
this result as tentative. Keeping this caveat in mind, we see a tendency 
that SNe Ia with a blueshift/redshift in the nebular lines show 
a higher/lower peak luminosity than expected by the Phillips relation. 
There is an average difference of $\sim 0.25$ mag in the 
peak magnitudes between SNe Ia showing a blueshift and those showing a redshift. 

Calculating the light curves based on the geometry derived by M10a for 
the relatively faint SN Ia 2003hv, 
we have found that SNe Ia viewed from the offset direction 
should have a peak brightness larger than the mean, and 
those viewed from the opposite direction should be 
fainter. The difference between these two extreme observer directions 
is $\sim 0.7$ mag in our fiducial model A0.3.\footnote[13]{Including 
the effect that the light curve shape is different for different viewing directions.}  
This behavior is consistent 
with the observational data. Also, the averaged difference between 
SNe showing blueshifts and redshifts is $\sim 0.4$ mag in this model, 
enough to explain the observed value ($\sim 0.25$ mag). 

The comparison between SNe Ia with normal/large peak luminosity and 
the models is less straightforward, since the geometry of such explosions 
has not been directly constructed from observations. 
Assuming that the degree of the offset is similar 
to that of SN 2003hv, we expect that the viewing angle 
effect is less pronounced for brighter SNe Ia. 
This is consistent with the behavior we have found in the data. 
Further variation is expected 
since the degree of the offset could be different for SNe with different 
$\Delta m_{15} (B)$ (e.g., Kasen et al. 2009; M10c). 
We therefore suggest that the residual could be explained 
by a combination of the configuration [e.g., 
the relative contribution between the ECAP/HD and LD zones, 
which may well be expressed by one parameter, i.e., $\Delta m_{15} (B)$] 
and the viewing angle. Thus, we do not expect a single straight 
line to give a perfect fit to the residual vs. late-time velocity plot. 
This may be one reason why 
the correlation in Fig.~8 is not too strong. 

\subsection{Future Perspectives and Cosmological Applications}

Our results could be further tested by polarization measurements. 
SNe Ia generally show small continuum polarization around maximum brightness, 
indicating that they are more or less spherical (e.g., Wang et al.\ 1996). However, 
early-phase polarization measurements mainly probe a region near the surface of the 
SN Ia ejecta, which we also suggest to be nearly spherical. Therefore, the small 
polarization is likely not a strong argument against our present interpretation. 
The low continuum polarization is sometimes translated into a small deviation 
of the photosphere from spherical symmetry, and hence a small dependence 
of the brightness on the viewing angle (e.g., H\"oflich et al.\ 2010). 
However, this statement depends strongly on the assumed geometry. 
For example, a continuum polarization of $\lsim 0.3\%$, as is usually found in SNe Ia, 
implies an axis ratio of less than 10\% for an ellipsoidal photosphere (H\"oflich\ 1991). 
However, for a one-sided distribution of $^{56}$Ni, as suggested in 
the present work, the expected continuum polarization is 
generally much smaller than for an ellipsoid (Kasen \& Plewa\ 2007) despite 
a larger expected variation in the angle-dependent brightness than for the ellipsoidal case 
(e.g., Sim et al.\ 2007; Kasen et al.\ 2009; this work). 

There is a correlation between the velocity gradient and the Si II line 
polarization (Leonard et al. 2005; Chornock \& Filippenko 2008; Maund et al. 2010). 
This may indicate that the viewing direction is indeed 
controlling the line polarization level, 
if the velocity gradient is determined by the viewing angle effect (M10b). 
Further study on the polarization of SNe Ia both in observations (e.g., Wang et al.\ 2007) 
and in theory (e.g., H\"oflich\ 1991; Kasen \& Plewa\ 2007) should provide 
a good test for the geometry of SNe Ia, and thus for the results in the present work. 

A problem in our analysis, especially for the residual issue, is the small sample size. 
The uncertainty in the distance measurement is also critical in our investigation of 
the viewing-angle effect on the luminosity residual. 
We suggest the following strategies 
to decrease the uncertainty in the distance estimate: 
(1) obtain late-time spectra for SNe~Ia at redshift $z \gsim 0.02$. 
At $z \sim 0.02$, the $V$-band magnitude of typical SNe Ia at $\sim 200$ days after 
maximum brightness is $\sim 21 - 22$ mag (depending on the extinction). 
Spectroscopy is thus possible with $6 - 8$m class telescopes, and 
(2) obtain comprehensive photometry during both the peak and tail phases. 
The residual arising from the viewing angle can in principle be seen 
in the peak-to-tail luminosity ratio, since the effect of the viewing angle 
vanishes at late-phases (e.g., Maeda et al.\ 2006b). This may provide 
a distance-independent measurement of the residuals caused by the viewing angle effect. 
Such observations are less demanding than spectroscopy, and thus can reach 
to larger redshift. 

One of the current limitations in 
SN Ia cosmology is the fact that a dispersion at the level of 
$\sim 0.15$ mag remains after the standardization of the peak luminosity 
with existing relations, and 
that the physical origin of the residual has not been identified. 
The viewing angle effect may account for 
part of the dispersion of the SN Ia luminosity 
calibration, although the presently available sample does not 
allow us to quantify how much improvement can be achieved by taking 
this effect into account. 
However, the effect of the random viewing angle 
enters in the SN Ia luminosity calibration as a source of a statistic error. 
Thus, increasing the number of SNe Ia for cosmology should effectively 
reduce this error in estimating cosmological parameters. 

In addition, it may be worthwhile looking into the frequency 
distribution of the residuals. Although we expect that the statistical error is introduced by 
random viewing angles, there is no reason to expect it to obey a Gaussian distribution. 
Investigating a non-Gaussian component in the scatter of the Hubble diagram may 
provide additional insight. For example, a larger SN Ia 
sample may allow us to estimate the non-Gaussian contribution 
to the statistical error, which will then provide a quantitative estimate of the 
viewing angle effect independent of uncertainties in the distance measurements 
for individual SNe. 

The result of the present work may shed light on how to develop 
a more accurate SN Ia cosmology than currently employed. 
The correlation between the color near maximum and $v_{\rm neb}$ 
enables us to discriminate the intrinsic color and the host extinction when late-time 
spectra are available. This opens up a possibility of studying the intrinsic color and 
the host extinction, including $R_{V}$, in detail for nearby SNe up to a redshift 
of $\sim 0.02$ for which late-time spectroscopy is possible. 
This will hopefully provide information about how to distinguish the 
intrinsic color and host extinction, the information applicable to high-redshift SNe. 
The relation between the 
velocity gradient and the nebular emission line shift (M10b) suggests that one could 
use the velocity gradient (accessible to higher redshift SNe) instead of the latter, in the 
color and luminosity calibrations (see also Foley \& Kasen 2010). 
Indeed, it has been argued that LVG and HVG SNe should be treated differently 
in the luminosity calibration (e.g., Wang et al. 2009b). Further investigating relations 
among these observable (e.g., Fig.~5), as well as finding other observables 
which correlate with $v_{\rm neb}$, 
could be useful in improving the color and luminosity estimates also for high-$z$ SNe~Ia.

\section*{Acknowledgments} 
The authors thank the anonymous referee for many constructive comments. 
A special thanks to the staff of Gemini South for their efforts in obtaining data and providing
user support. 
This research is supported by World Premier International Research Center
Initiative (WPI Initiative), MEXT, Japan. 
K. M. acknowledges financial support by Grant-in-Aid for
Scientific Research for young Scientists (20840007). 
The work has been partly done during the visit of K. M. to MPA 
supported by the Max-Planck Society and to Stockholm Observatory 
supported by the Oskar Klein Centre and the Scandinavia-Japan Sasakawa Foundation. 
S.T. acknowledges support by the Transregional Collaborative Research Centre TRR 33 
`The Dark Universe' of the German Research Foundation (DFG). 
M.S.  is supported by the National Science Foundation (NSF) under grant AST--0306969.
J.S. is a Royal Swedish Academy of Sciences Research Fellow supported by a grant from the Knut and Alice Wallenberg Foundation. 
S.B. acknowledges partial support from ASI contracts `COFIS'.
M.H. acknowledges support by FONDECYT Regular 1060808, 
Centro de Astrofisica FONDAP 15010003, 
Centro BASAL CATA (PFB 06), and the Millennium Center for Supernova Science 
(P06-045-F). 
The Dark Cosmology Centre is funded by the Danish 
National Research Foundation.
This research made use of the {\it SUSPECT} (the online Supernova Spectrum Archive), 
at the Department of Physics and Astronomy, University of Oklahoma, 
and the HyperLeda database (http://leda.univ-lyon1.fr).


\appendix

\section{Extinctions as compared to Other Estimates}

\begin{table*}
\centering
 \begin{minipage}{140mm}
 \caption{Extinction from the literature}
 \label{tab:mvvneb}
 \begin{tabular}{@{}lccc@{}}
 \hline 
SN & $A_{V}$& Method & References\footnote{Refs. 
A04. Altavilla et al. (2004); 
K00. Krisciunas et al. (2000); 
K04. Krisciunas et al. (2004); 
K07. Krisciunas et al. (2007); 
L01. Li et al. (2001); 
L09. Leloudas et al. (2009); 
P07. Pastorello et al. (2007a); 
P08. Pignata et al. (2008); 
Ph99. Phillips et al. (1999); 
S07. Stanishev et al. (2007); 
S10a. Stritzinger et al. (2010); 
S10b. Stritzinger et al. (in prep.);  
W08. Wang et al. (2008); 
W09a. Wang et al. (2009a); 
W09b. Wang et al. (2009b)} \\
  & (mag) & & \\
\hline
1990N & $0.16$ &       & Ph99\\
1994D & $0.0$ &         & Ph99\\
1997bp & $0.46$ &      & A04\\
1998aq & $0.0$ &        & W09b\\
1998bu & $1.06$ & NIR & K00\\
2000cx & $0.0$ &        & L01\\
2001el & $0.47$ & NIR & K07\\
2002bo & $0.51$ & NIR & K04 \\
2002dj & $0.0$ & NIR    & P08\\
2002er & $0.40$ &        & W09b\\
2003du & $0.0$ & NIR    & S07\\
2003hv & $0.0$ &         & L09\\
2004dt & $0.19$ &        & W09b\\
2004eo & $0.0$ & NIR   & P07\\
2005cf & $0.18$ & NIR & W09a\\
2006X & $2.1$ & NIR  & W08 \\
2006dd & $0.12$ &   & S10a\\
2007on & $0.25$ &     & S10b\\
2007sr & $0.13$ &      & S10b\\
2009ab & $0.076$ &    & S10b\\
\hline 
\end{tabular}
\end{minipage}
\end{table*}

\begin{figure}
   \centering
   \includegraphics[width=0.45\textwidth]{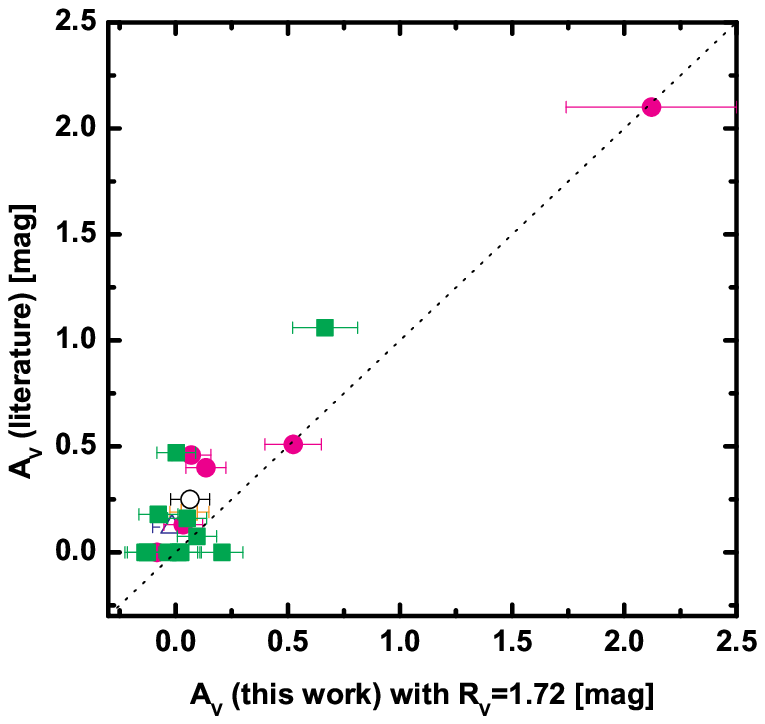}
   \caption{Comparison of the extinction. The horizontal axis denotes the values of 
$A_{V}$ derived in this study (Tab.~2), while the vertical axis provides values from 
literatures (Tab. A1). The symbols and color coding indicate the velocity gradient near 
maximum brightness (see Fig. 4 caption). 
}
   \label{figa1}
\end{figure}

In \S 4, we compared the color excesses we derived (with $v_{\rm neb}$) 
to those estimated by Wang et al. (2009b). 
As a further, additional test, we also check how our extinction values compare to 
those previously estimated and presented in the literature by different methods. 
Assuming $R_{V} = 1.72$ (\S 1, \S 5), we convert $E(B-V)$ to $A_{V}$. 
Figure A1 shows a comparison between $A_{V}$ as derived in this study 
and the values from the literature. 
The literature values were selected as follows: 
if optical minus near-IR  (i.e. $V - K$, $V-J$, $V-H$) color curves are available, 
the template color curves of Krisciunas et al. (2000, 2001, 2009) were
used to derive $A_{\rm V}$ 
with the infrared extinction law given by Rieke \& Lebofsky (1985). 
A similar estimate is also possible using the near-IR 
spectral energy distribution (SED) at a single epoch [e.g., Wang et al.\ (2008), 
who used the extinction law of Cardelli, Clayton, \& Mathis (1989) constructed from the 
data provided by Rieke \& Lebofsky (1985)]. 
For SN~2001el, the value of $A_{V}$ was taken from Krisciunas et al. (2007) 
who used SN~2004S as a template because of the similarity between these two events. 
When these near-IR measurements were available, we adopted $A_{V}$ based on these estimates. 
When near-IR measurements were not available, we converted 
$E(B-V)$ color excess estimates {\em from the literature} to $A_{V}$ using $R_{V} = 1.72$.  
In the cases of  SNe~2007sr  (Schweizer et al.\ 2008), 2007on and 2009ab (Stritzinger et al.\ in prep.), 
$E(B-V)$ was obtained using the template light curve fitter SNooPy (Burns et al. 2011). 
The values thus compiled are listed in Table A1. 

From this comparison, we find the similar results to those obtained by the comparison to 
$E(B-V)$ derived by Wang et al. (2009b). Our estimates tend to be smaller, and 
even if we assume $R_{V} \sim 3$ to convert our $E (B-V)$ estimate to $A_{\rm V}$, 
the extinctions we derive are mostly consistent with those derived by the other method 
(except for the heavily reddened SN 2006X).

\section{Uncertainties in the Luminosity Calibration} 

\begin{table*}
\centering
 \begin{minipage}{140mm}
 \caption{Relations between $v_{\rm neb}$ and $dm$.}
 \label{tab:mvvneb}
 \begin{tabular}{@{}lcrrrc@{}}
 \hline 
Description & N\footnote{The number of SNe.} & 
$\alpha$\footnote{The best fit using the relation 
$dm = \alpha \ (v_{\rm neb})/1000 \ {\rm km} \ {\rm s}^{-1}) + \beta$ mag. 
The errors for $\alpha$ and $\beta$ are $1\sigma$ uncertainties.} & 
$\beta$ & 
$P$\footnote{Probability that the distribution arises from a non-correlation.} & 
significance\\
\hline
Distance\footnote{Obtained by changing the distances to SNe 1990N and 2006dd (\S B1).} 
             & 20 & $0.092 \pm 0..052$ & $0.070 \pm 0.077$ & 0.038 & 
$1.8\sigma$\\
STS\footnote{Obtained by changing the distances to the STS scale (\S B1; Table B2).}
             & 20 & $0.11 \pm 0.050$ & $-0.013 \pm 0.076$ & 0.016 & 
$2.1\sigma$\\
$R_{V}$\footnote{With $R_{V} = 3.1$ (\S B2).}
             & 20 & $0.072 \pm 0.055$ & $0.040 \pm 0.082$ & 0.099 & 
$1.3\sigma$\\
$A_{V}$ (literature)\footnote{$A_{V}$ adopted from the literature (\S B2).} 
             & 20 & $0.050 \pm 0.049$ & $-0.064 \pm 0.074$ & 0.16 & 
$1.0\sigma$\\
$\bar M_{V} (\Delta m_{15} (B))$\footnote{The reference magnitude changed to that of Phillips et al. (1999) (\S B3). } 
             & 20 & $0.081 \pm 0.051$ & $0.071 \pm 0.077$ & 0.059 & 
$1.6\sigma$\\
\hline
\end{tabular}
\end{minipage}
\end{table*}

Compared to the color calibration (\S 4), 
our analysis on the luminosity residual (\S 6) 
suffers from various sources of uncertainties. 
These are independent from $v_{\rm neb}$, and so 
it is unlikely that any correlation we investigate in the present study 
is artificially introduced by these uncertainties in a systematic way. 
However, our present sample is still small and thus statistical errors 
due to these uncertainties can still be important. 
In this section, we investigate how our results in \S 6 are 
affected by changing the procedures to estimate the distance, reddening, and 
standardized luminosity (\S 5).  
Results are summarized in Table B1. Each item in the table is 
explained in the following.

\subsection{Distances}

\begin{figure}
   \centering
   \includegraphics[width=0.45\textwidth]{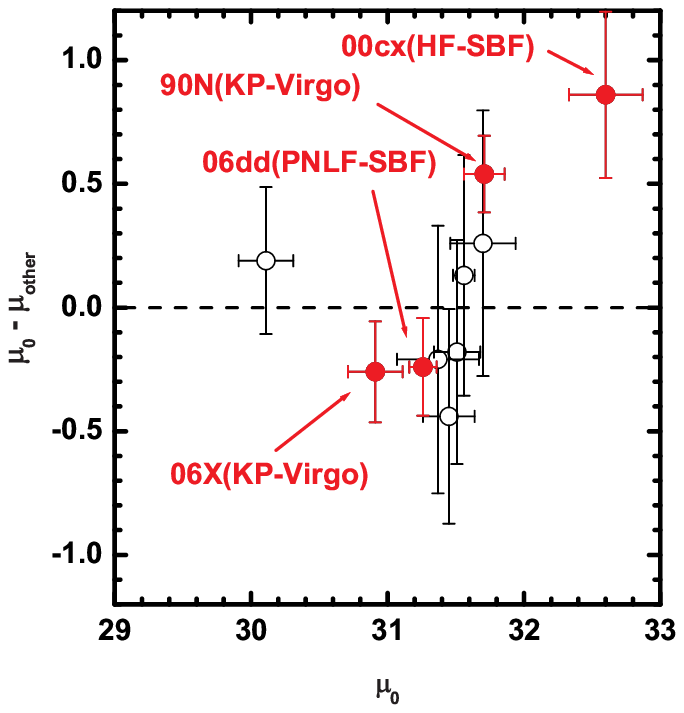}
   \caption{Comparison of the distance we adopted ($\mu_{0}$) and 
that from a different method ($\mu_{\rm other}$) for SNe 
with several independent distance measurements available. 
SNe discussed in \S B1 are marked by red-filled circles and 
indicated by SN ($\mu_0 - \mu_{\rm other}$). 
}
   \label{figb1}
\end{figure}

To examine the uncertainty in the distances, we first compare 
the distances obtained by different methods, when available, in Fig.~B1. 
Hubble flow (HF) distances with a recession velocity  
below $1,000$ km s$^{-1}$ are not considered as alternatives here.  
Different measurements typically agree with each other to within the errors 
(albeit these errors are sometimes 
quite large, when only the HF distances are available as the `other' measurement). 
However, there are several exceptions. 
NGC 4639 (SN 1990N) and NGC 4321 (SN 2006X) are (possible) members of 
the Virgo cluster. If we adopt the distance modulus $\mu = 31.17 \pm 0.04$ mag 
for the Virgo cluster 
(Kelson et al.\ 2000), it differs significantly from the Cepheid distance to NGC 4639 
we adopted ($\mu$ (KP) $= 31.71 \pm 0.15$ mag). 
We note, however, that Riess et al. (2009) revised the Cepheid distance to NGC 4639 to 
be $\sim 31.48 \pm 0.13$ mag. Although there is still a discrepancy between the distances 
to NGC 4639 and the Virgo cluster, this new measurement makes the agreement better. 
If we adopt the value of Riess et al. (2009), SN 1990N 
should be fainter by 0.23 mag than our fiducial estimate. 
We did not adopt this value in the main text, to avoid possible systematic errors 
in different Cepheid measurements. The distance to NGC 524, 
the host of SN 2000cx is $31.74 \pm 0.20$ mag (SBF) or $32.60 \pm 0.27$ mag (HF), 
which clearly do not agree with each other. 
Although we usually adopt the SBF distance as the better estimate, we note that 
another supernova, SN 2008Q, appeared in the same galaxy and 
{\it both} of these SNe would be peculiar outliers in this case. 
Thus, we adopted the HF distance to SN 2000cx (\S 3). Finally, the PNLF distance 
($\mu = 31.26 \pm 0.1$ mag) to NGC 1316 (SN 2006dd) is smaller than the SBF distance 
($\mu = 31.50 \pm 0.17$), and Stritzinger et al. (2010) argued that the former is 
more accurate (thus adopted in the main text). 

Following the above inspection, we change the distance to SN 1990N (from the original KP 
value to that derived by Riess et al.\ 2009), and that to SN 2006dd (from the PNLF to the SBF), 
and repeat the same analysis as we did in Fig.~8. We thereby obtain 
a chance probability of 
$P = 0.038$ ($1.8\sigma$) (Table B1).

\begin{table}
\centering
 \begin{minipage}{140mm}
 \caption{SNe Ia sample on the STS distance scale.}
 \label{tab:sample}
 \begin{tabular}{@{}lll@{}}
 \hline
 SN & Host & $\mu$\\
      &        & (mag)\\ 
\hline
1990N & NGC 4639 & $32.20 \pm 0.09$ (STS)\\
1994D & NGC 4526 & $31.18 \pm 0.20$ (SBF)\\
1998aq & NGC 3982 & $31.87 \pm 0.15$ (STS)\\
1998bu & NGC 3368 & $30.34 \pm 0.11$ (STS)\\
2002bo & NGC 3190 & $31.90 \pm 0.24$ (SBF)\\
2003hv & NGC 1201 & $31.57 \pm 0.3$ (SBF)\\
2006X & NGC 4321 & $31.18 \pm 0.05$ (STS)\\
2007on & NGC 1404 & $31.65 \pm 0.19$ (SBF)\\
\hline
\end{tabular}
\end{minipage}
\end{table}

\begin{figure}
   \centering
   \includegraphics[width=0.45\textwidth]{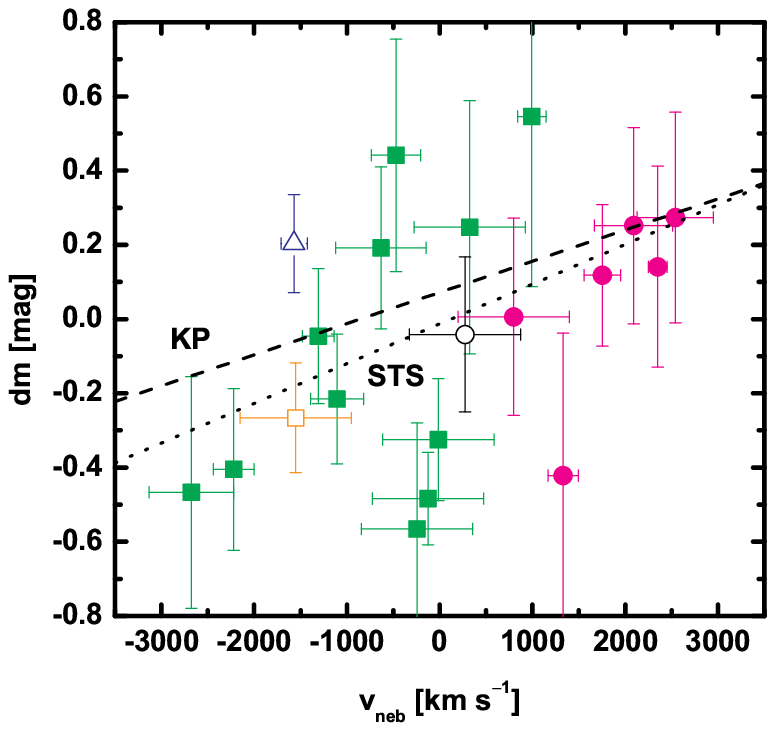}
   \caption{The same as Fig.~8, but using the STS measurements  
for the Cepheid distances and a revised 
zero point for the SBF distances. See the caption of Fig.~4 for 
the meaning of the different symbols. The dotted line is 
the best-fit line for the STS distances, while the dashed line is 
that for the KP distances. 
}
   \label{figb2}
\end{figure}

As another test, we explore how our results are affected if we adopt the STS 
Cepheid distances (\S 5). For 4 SNe Ia with a Cepheid distance available, 
there are also measurements by the STS group (Saha et al.\ 2006). 
We also change the SBF measurements to be consistent with 
the STS measurements. The SBF values used in the main text 
are calibrated with the KP Cepheid distance zero-point. 
Since the STS values are on average larger than the KP values by $\sim 0.2$ 
mag, we add $0.2$ mag for the SBF distances. 
Table B2 lists the STS distances thus derived for SNe Ia. 
For SNe Ia missing in Table B2, the same distances as in Table 1 are used. 
Figure B2 shows the $v_{\rm neb}$ vs. $dm$ diagram with the STS distance scale. 
For 20 SNe Ia, 
we obtained 
a chance probability of $P = 0.016$ ($2.1\sigma$). 
The slope of the linear fit is steeper than that derived with the KP distance scale, 
although they are consistent within the errors of the fits. The zero-point is 
displaced by $\sim 0.1$ mag, reflecting the difference between 
the STS and KP distance scales.

\subsection{Extinction}

Another large uncertainty comes from the estimate of the extinction 
within the host galaxies. 
To check the uncertainty, we change the value of $R_{V}$ to $3.1$, the standard 
value for Galactic extinction. 
Following \S 4 and F10, $R_{V}$ for SN 2006X is left unchanged. 
While it has been argued that the typical Galactic value 
is not necessarily applicable for the host galaxies of SNe Ia (e.g., Folatelli et al.\ 2010; 
Hicken et al.\ 2009ab; Wang et al.\ 2009b; Yasuda \& Fukugita\ 2010), 
our analysis in \S 4 suggests that an $R_{V}$ close to the typical Galactic value can be 
acceptable for mildly extinguished SNe, 
once the intrinsic color variation due to the viewing direction 
is taken into account. 
Figure B3 shows the $v_{\rm neb} - dm$ diagram with $R_{V} = 3.1$. 
We obtain linear fit parameters as follows; 
$P = 0.099$ ($1.3\sigma$). 
The best fit lines for different $R_{V}$ (1.72 and 3.1) 
are consistent to each other within the fitting errors. The significance of the fit is 
slightly weaker for $R_{V} = 3.1$ than in our fiducial case. 
It is, however, a reasonably large range of $R_{V}$ that we investigate. 

\begin{figure}
   \centering
   \includegraphics[width=0.45\textwidth]{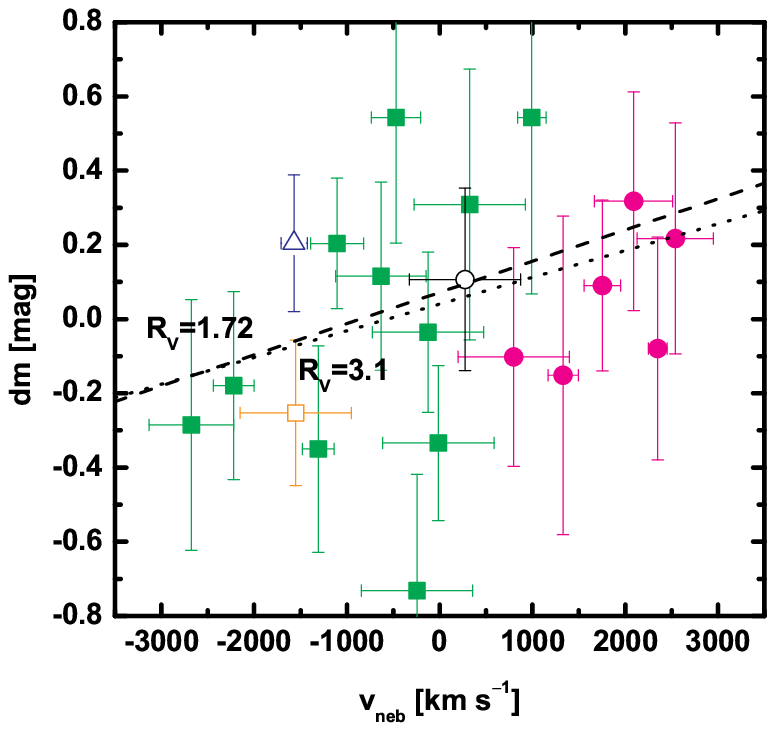}
   \caption{The same as Fig.~8, but using $R_{V} = 3.1$ for 
the extinction. See the caption of Fig.~4 for 
the meaning of the different symbols. The dotted line is 
the best-fit line for the data points with $R_{V} = 3.1$, while the dashed line is 
for the fiducial value of $R_{V}  = 1.72$. 
}
   \label{figb3}
\end{figure}

\begin{figure}
   \centering
   \includegraphics[width=0.45\textwidth]{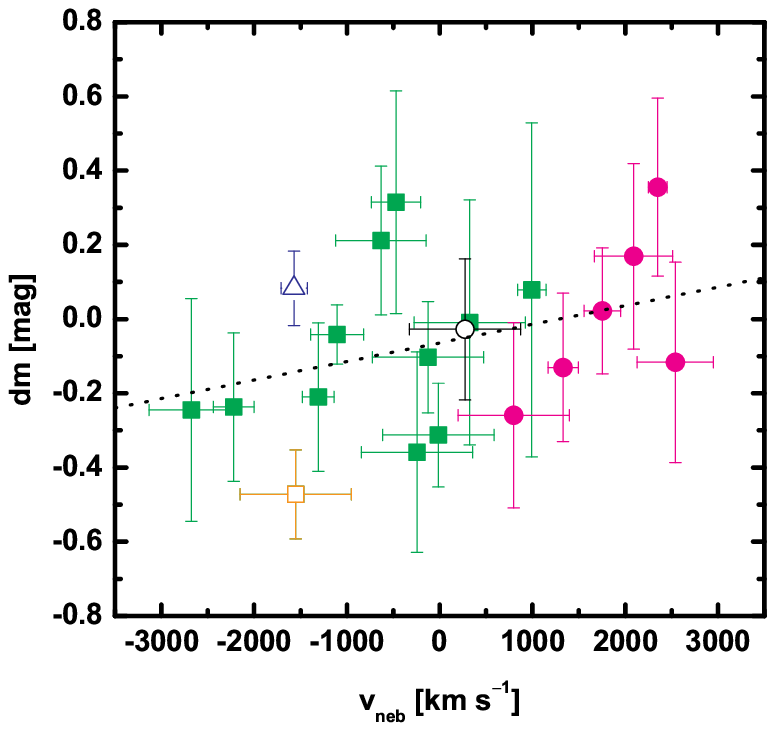}
   \caption{The same as Fig.~8, but $A_{V}$ replaced by the values in the literature 
(Table A1). The best-fit line is shown by the dotted line. 
}
   \label{figb4}
\end{figure}

Next, we explore whether our use of the relation between $v_{\rm neb}$ and the color 
to derive the extinction might affect the results. For this purpose, we replace the 
host galaxy extinction by the values given in the literature, and repeat the same analysis. 
For the literature values, see Table A1 and the related discussion in the main text. 
The result is shown in Figure B4. 
The fitting result is $P = 0.16$ ($1.0\sigma$) (Table B1). 
Although there is still a correlation, 
it is weaker than if the $v_{\rm neb}$ -- color 
relation derived in this paper is adopted ($1.6\sigma$).

\subsection{Standardized Luminosities}

\begin{figure}
   \centering
   \includegraphics[width=0.45\textwidth]{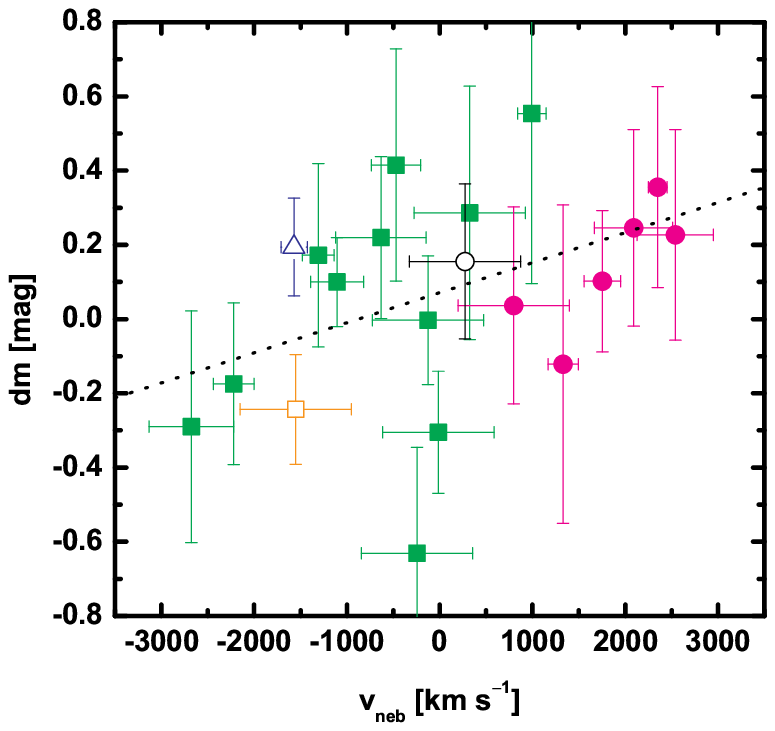}
   \caption{The same as Fig.~8, but using the original Phillips 
relation (Phillips et al. 1999) for the reference magnitude 
($\bar M_{V} (\Delta m_{15} (B))$). See the caption of Fig.~4 for 
the meaning of the different symbols. The dotted line is 
the best-fit line for this data set. 
}
   \label{figb4}
\end{figure}

By definition, $dm$ depends on the standardized luminosity, i.e., 
$\bar M_{V} (\Delta m_{15} (B))$. We have used an 
updated relation given by F10. 
To check the related uncertainty, we replace 
$\bar M_{V} (\Delta m_{15} (B))$ 
by the original version of the Phillips relation including the second order term in 
$\Delta m_{15} (B)$ (Phillips et al. 1999) and repeat the same analysis. 
$\bar M_{V} (\Delta m_{15} (B) = 1.1)$ is set to be $-19.12$ mag, 
to be consistent with Folatelli et al. (2010). 
Figure B5 shows the resulting $v_{\rm neb} - dm$ diagram. 
We obtain $P = 0.059$ ($1.6\sigma$). 

\section{Monte-Carlo Simulation for Estimating the Chance Probability}

To estimate the chance probability $P$ that a distribution arises from a non-correlation, 
we have used the following method based on Monte-Carlo simulations in \S 6.:
we performed a linear regression to a set of 
variables ($X_i \pm \sigma_{X_i}, Y_i \pm \sigma_{Y_i}$) (where $i$ spans the numbers in 
the SN sample). 
First, we produced $10^{5}$ test distributions obtaining ($X'_{i, k}, Y'_{i, k}$) 
(with $k$ spanning from $1$ to $10^{5}$), 
where a Gaussian distribution is assumed for the variation 
in $X_i$ and $Y_i$ with the associated errors $\sigma_{X_i}$ and $\sigma_{Y_i}$. 
Here, $\sigma_{Y_i}$ includes both the extinction and distance uncertainties. 
For each $k$-th test distribution, we performed a linear regression fitting 
assuming the functional form $Y' = \alpha_k X' + \beta_k$. 
Here we used only the error associated with 
the extinction as a weight in the $\chi^2$ fitting for each distribution, 
because the `relative' errors between different distance measurements are 
difficult to quantify. 
We thereby obtained $\alpha_k = \alpha_{k, 0} \pm \sigma_{\alpha_k}$ 
and $\beta_k = \beta_{k, 0} \pm \sigma_{\beta_k}$ by the linear regression 
for each $k$-th test distribution. This results in a probability 
distribution of the fitting coefficients, $\alpha$ and $\beta$, 
by convolving $10^{5}$ Gaussian distributions for $\alpha_{k}$ and $\beta_{k}$. 
The final fitting result is then obtained by 
fitting Gaussian profiles to the probability distribution of $\alpha$ and $\beta$. 
To estimate the chance probability $P$ that the distribution ($X \pm \sigma_{X}$, 
$Y \pm \sigma_{Y}$) would arise from a non-correlation, we counted 
the probability for $\alpha \le 0$ (if the mean value of $\alpha$ is positive).


\bsp

\label{lastpage}

\end{document}